\documentclass[aps,pra,preprint]{revtex4-1}
\usepackage{amsmath,amsfonts,amsthm} 
\usepackage{graphicx}
\usepackage{xcolor}
\usepackage{physics}
\usepackage{hyperref}
\usepackage{bm}
\begin{document}

\title{Intrinsic decoherence and recurrences in a large ferromagnetic $F = 1$ spinor Bose-Einstein condensate}

\author{J.C. Sandoval-Santana$^1$, R. Zamora-Zamora$^2$, R. Paredes$^3$, and V. Romero-Roch\'{\i}n$^3$} 

\affiliation{$^{1}$ Departamento de Ciencias B\'asicas,
Universidad Aut\'onoma Metropolitana Azcapotzalco,
Av. San Pablo 180, Col. Reynosa Tamaulipas,  02200 Ciudad de M\'exico, Mexico}
\affiliation{$^{2}$ Department of Applied Physics, QCD Labs, COMP Centre of Excellence, Aalto University, PO Box 13500, Aalto, FI-00076, Finland.}
\affiliation{$^{3}$ Instituto de F\'{\i}sica, Universidad Nacional Aut\'onoma de M\'exico,
Apartado Postal 20-364, 01000 Ciudad de M\'exico, Mexico.}

\begin{abstract} 
Decoherence with recurrences appear in the dynamics of the one-body density matrix of an $F = 1$ spinor Bose-Einstein condensate, initially prepared in coherent states, in the presence of an external uniform magnetic field and within the single mode approximation. The phenomenon  emerges as a many-body effect of the interplay of the quadratic Zeeman effect, that breaks the rotational symmetry, and the spin-spin interactions. By performing full quantum diagonalizations very accurate time evolution of large condensates are analyzed, leading to heuristic analytic expressions for the time dependence of the one-body density matrix, in the weak and strong interacting regimes, for initial coherent states. We are able to find accurate analytical expressions for both the decoherence and the recurrence times, in terms of the number of atoms and strength parameters, that show remarkable differences depending on the strength of the spin-spin interactions. The features of the stationary states in both regimes is also investigated. We discuss the nature of these limits in the light of the thermodynamic limit.
\end{abstract}

\maketitle

\section{Introduction}

The observation of stationarity in quantum systems relies on the existence of pair wise collisions in large conglomerates of atoms, either in an isolated environment or in contact with a larger environment \cite{Redfield1965theory,Lindblad,van1992stochastic,van1995soluble,van2004new,Scully,Zurek,Caldeira-Leggett,Silbey,Romero-Rochin}. While in the former case the process through which the system reaches the stationary state may be termed {\it intrinsic} decoherence \cite{Caballero-Benitez,Camacho-Guardian}, in the later case it is simply known as decoherence. In reality, since one can consider the system $(S)$ under study and the environment $(R)$ as a composite {\it closed} system $(S+R)$, their time evolution is always unitary and the observation of decoherence always refers to {\it reduced} quantities, namely, observables of much fewer degrees of freedom of the total ones of the isolated system, say $N_S \ll N_R$, with $N$ the number of degrees of freedom. Therefore, when studying a system in contact with a reservoir, one deals with the reduced density matrix of the former, $\rho_S$, integrating out the environment. This leads typically to master equations for the reduced density matrix of the system that has already ingrained the decoherence effects induced by the interaction with the environment \cite{Lindblad,van2004new,Scully,Caldeira-Leggett,Silbey,Romero-Rochin}. In general, decoherence and recurrence phenomena have been widely studied both, experimentally and theoretically. Within the experimental side some examples are the dephasing in interference fringes measured in condensates, the decay of laser induced polarization in spectroscopic experiments \cite{Engel,Collini}, the amplitude damping in qudit photonic states \cite{Marques}, and the decoherence induced in single molecule junctions \cite{Ballmann} among others. Measurements of purity have also been used to search for quantum coherence loss \cite{Gu}. 

On the other hand, if the system under study is a large one, $N \gg 1$, but isolated from any environment, even though the evolution is always unitary, the system does relax or decohere to a stationary state \cite{LL} within a quantifiable decoherence time, in the sense that expectation values of one-body observables, such as temperature, magnetization and, say, few-body correlations, behave as if the full system had relaxed to a stationary state for times longer than the decoherence time. 
If the system is finite, there appear recurrences or revivals of states near the initial one, but in typical systems the recurrence time may grow  with no bound in the thermodynamic limit $N \to \infty$. The behavior of those few body properties can be directly studied solely with their corresponding reduced few-body density matrices, whose time evolution is no longer unitary within their reduced Hilbert space. It is of interest for our purposes that the experiments in ultracold gases showing Bose-Einstein condensation \cite{Anderson198,Ketterle1995,Jin2004,Vinit2017} are the closest in dealing with true isolated systems. Yet, these gases, due to atomic collisions, do relax to equilibrium and, when in the presence of external magnetic fields do show decoherence phenomena \cite{PRA2019Mitchell,PRL2020Mitchell}. The study of the magnetization of the latter case is the subject of the present article. Hence, to be specific, we call intrinsic decoherence to that of few-body properties, in an otherwise very large isolated system. Recurrences are not observed in the ultracold gases since unwanted processes, such as three-body collisions, make the thermal state unstable in a relatively short time \cite{Ketterle1999}. An interesting question concerns the nature of the stationary state and the enquiry if it is a thermal state or not. In this regard the essence of the Eigenstate Thermal Hypothesis (ETH) \cite{Deutsch,Srednicki1994,Rigol,Reimann_2010,Kaufman794} is to establish that the thermodynamic properties of few body observables are contained in the eigenstate closest to the equilibrium average energy. Hence, a simple test is to compare the few-body density matrices of the stationary state of the actual unitarily evolving state with those of the energy eigenstate with an energy similar to the mean of the state evolving in time. 
Deviations from this typical behavior, such as the many-body localization phenomenon \cite{choi1547}, are also of current interest. In the light of these observations, we advanced here that depending on the two-body interaction strength in an $F = 1$ spinor Bose-Einstein (SBEC) condensate in the presence of external homogeneous magnetic fields, we do observe intrinsic decoherence of the magnetization but with different features leading to both, typical and non-typical stationary states.

To be precise, within a full quantum scheme, we study here the phenomenon of intrinsic decoherence as well as the appearance of recurrences, in the time dynamics of an spinor $F = 1$ Bose-Einstein condensate (SBEC) composed of a mixture of three different hyperfine spin components, in the presence of a uniform magnetic field, starting in a well defined coherent state. Depending on the sign of the spin-mixing interaction strength, the atomic cloud can be polar if positive, such as a gas of  $^{23}$Na atoms, or ferromagnetic if negative, as in a gas of $^{87}$Rb atoms \cite{ho1998spinor,Machida1998}. We shall study here the ferromagnetic case only since the condensate acquires a macroscopic spin texture that makes it behave as a ``giant'' spin.
It is worth mentioning that the system we address is very similar to the recent experimental investigation on the spin dynamics of an $F=1$ $^{87}$Rb spinor macroscopic condensate, where use of the SBEC as a sensible magnetometer is explored  \cite{PRL2020Mitchell}. The present model has also been used to study quantum phase transitions in space \cite{Damski}. The dynamics of the spin mixture is followed in the presence of the external magnetic field that gives rise to linear and quadratic Zeeman contributions, with strengths $p$ and $q$, which together with the interaction term, of strength $\eta$, give rise to the phenomena here analyzed.

Since we are able to very accurately diagonalize the Hamiltonian of the system up to $N \sim 10^4$ atoms, we can study the dynamics of any initial state and then calculate reduced density matrices of few bodies. Here, we analyze the one-body density matrix for a full family of coherent states, as one expects them to be the most resilient to the spin interaction, clearly showing Larmor-like oscillations in the expectation value of the measurable spin (or magnetization) one-body observable. We point out that coherent states are usually expected to yield the closest to quasi-classical dynamics of the magnetization, in a mean-field fashion, with little or no decoherence \cite{kajtoch2016spin}. And indeed, if there is no quadratic Zeeman coupling,
 the coherent states show no signs of decoherence, but as soon as the full rotational symmetry is broken by the presence of such a coupling, this, in conjunction with the natural two-body collisions yield decoherence or collapse of the very definite Larmor oscillations generated by the presence of the external field. Due to the finite number of atoms and the relatively small Hilbert space size, the oscillations recur or revive after longer periods of time, to decohere again. After a long set of calculations we were able to find out three clear regimes depending on the value of the dimensionless parameter $N|\eta|/q$, a weak interacting one $N|\eta|/q \ll 1$, a strong one $N|\eta|/q \gg 1$ and a crossover $N|\eta|/q \sim 1$, in which, although all show decoherences and recurrences, their dependence on the parameters $q$ and $\eta$ and specially on $N$, as well as the nature of their corresponding stationary states, are very different. Also very importantly, after the mentioned numerical study, and with the insight of the analytic solution to the non-linear single mode model\cite{Imamoglu,plimak2006quantum}, we were able to heuristically deduce analytic expressions for the full one-body density matrix in the weak and strong interacting regimes. These expressions are certainly the leading order contributions of the full unknown analytical solution and show a remarkable agreement with the predictions of the full quantum numerical solution. We are thus able to provide analytical expressions for the recurrence and decoherence times in terms of the parameters $q$, $\eta$, $N$ and their dependence on the particular initial states. We also highlight the fact that the strong interaction regime shows a ``typical'' behavior of a macroscopic system since the decoherence and recurrence times show a expected dependence on $N$, in the sense that as $N \to \infty$, the recurrences time grows with no bound, thus making the quasistationary state closer to a true one. In addition, the nature of the stationary state behaves similarly to ETH as having the same reduced one-body density matrix as the eigenstate whose energy equals the gas average energy.

This manuscript is organized in 5 sections. In section 2 we introduce the model Hamiltonian that describes the spinor BEC within the single mode approximation (SMA) and discuss how we are able to accurately obtain its full quantum diagonalization. In section 3 we introduce the one-body density matrix and its time evolution for a family of coherent states in the ferromagnetic case, preparing the stage for section 4 that shows our main contributions regarding the discussion of decoherence recurrences in the weak and strong interacting regimes. Finally a discussion and a summary of this work is presented in section 5.

 \section{An $F = 1$ SBEC within the SMA approximation. Full quantum diagonalization.}

\begin{figure}[htbp]
\begin{center}
\includegraphics[width=.6\textwidth]{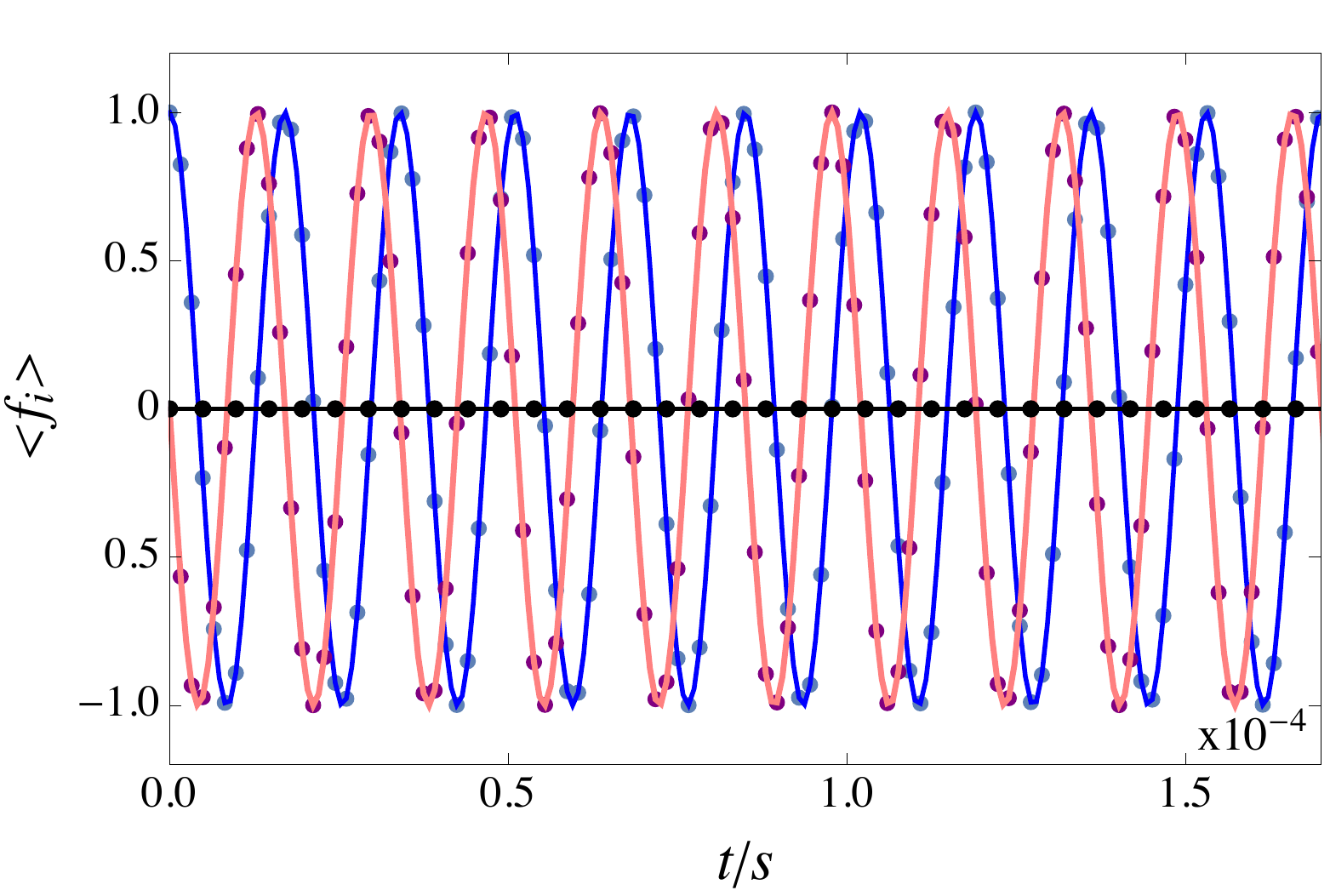}
\end{center}
\caption{Magnetization $\vec f$ as a function of time $t$. Comparison of a full 3D Gross-Pitaevskii (dotted line) versus SMA-GP calculations (continuous line), for a $^{87}Rb$ $F=1$ ferromagnetic SBEC. Blue and red lines are $x-$ and $y-$components, with black $z-$component. We use $^{87}Rb$ constants and experimentally accesible fields, $\tilde p = - 0.7 h$ MHz G$^{-1}$, $\tilde q = 72 h$ Hz G$^{-2}$, $c_0 = 50.2$ \AA, $c_2 = 50.9$ \AA$\>$  with a field $B_z = 84$ mG and for $N = 6.8 \times 10^4$ atoms.}
\label{figura0}
\end{figure}

The many-body  $F = 1$ SBEC Hamiltonian with linear and quadratic Zeeman couplings to an external homogeneous magnetic field $\vec B$, within the contact approximation, is
\begin{eqnarray}
{\cal H} &=& \int  d {\bf r} \> \left( \frac{\hbar^2}{2m} \nabla \hat \psi_\alpha^\dagger ({\bf r}) \cdot \nabla\hat \psi_\alpha ({\bf r}) + U({\bf r}) \hat  \psi_\alpha^\dagger ({\bf r})  \hat  \psi_\alpha({\bf r})+ \frac{c_0}{2} \hat  \psi_\alpha^\dagger ({\bf r})\hat \psi_\beta^\dagger ({\bf r}) \hat \psi_\beta ({\bf r}) \hat \psi_\alpha ({\bf r})    \right. \nonumber \\
&& + \frac{c_2}{2} \hat  \psi_\alpha^\dagger ({\bf r})\hat  \psi_{\alpha^\prime}^\dagger ({\bf r}) {\bf F}_{\alpha \beta} \cdot {\bf F}_{\alpha^\prime \beta^\prime} \hat \psi_\beta ({\bf r})  \hat \psi_{\beta^\prime} ({\bf r})  +   \tilde p \hat \psi_\alpha^\dagger ({\bf r}) \left[\vec B \cdot {\bf F}\right]_{\alpha \beta} \hat \psi_\beta ({\bf r}) \nonumber \\
&& \left. +  \tilde q  \hat \psi_\alpha^\dagger ({\bf r})  \left[\vec B\cdot {\bf F}\right]_{\alpha \alpha^\prime}  \left[\vec B \cdot {\bf F}\right]_{\alpha^\prime\beta} \hat  \psi_{\beta} ({\bf r})\right)\label{Htot}
\end{eqnarray}
where $\hat  \psi_\alpha ({\bf r})$ are annihilation operators of particles at ${\bf r}$ with spin $\alpha = -1,0,+1$ and ${\bf F}_{\alpha \beta}$ are the $F = 1$ angular momentum matrices. $c_0$ and $c_2$ are interaction coefficients proportional to the corresponding $s$-wave scattering lengths. If $c_0 > c_2$ the system is polar and for $c_0 > c_2$ ferromagnetic \cite{ho1998spinor}. $U({\bf r})$ is an external confining potential, typically harmonic. In general, the field operator is given by
\begin{equation}
\hat \psi_\alpha ({\bf r}) = \sum_{m} \phi_{m \alpha} (\vec r) \hat  b_{m\alpha}
\end{equation}
where $\phi_{m \alpha}(\vec r)$ are elements of a basis of the one-particle Hilbert space and $\hat b_{m\alpha}$ the corresponding creation operators. For ultracold gases, a usual approximation is to consider a self-consistent to be determined ground state wavefunction only $\Psi_{0\alpha}({\bf r},t)$, such that $\hat \psi_\alpha ({\bf r},t) \approx \Psi_{0\alpha}({\bf r},t) \hat  b_{0}$ , that is all spin states are in the same ground state. This leads to a set of three coupled Gross-Pitaevskii (GP) equations\cite{ho1998spinor,Machida1998}. These can be solved numerically and, as shown in Fig. \ref{figura0}, the solution for a {\it homogenous} magnetic field shows that the spatial part is unaffected by the presence of such a field, showing Larmor-like oscillations of the magnetization, see below. That is, the dynamics occurs only in the spin degrees of freedom and it is not transferred to spatial excitations such as phonons and vortices. This would not be so if the external field is inhomogeneous \cite{Zamora_Zamora_2018}. The previous observations indicates that a single-mode approximation can be made at the level of GP description (SMA-GP), that is, an assumption that the ground state wavefunction is the same for all spin components $\alpha$ . So, if we set $ \Psi_{0\alpha}({\bf r},t) = \psi_0(\vec r) \zeta_{\alpha}(t)$ with $\zeta_{\alpha}(t)$ a simple 3-component spinor, we can integrate out the spatial part and find very simple dynamical equations, called SMA-GP (not shown here), for the time evolution of $\zeta(t)$. The result of those is shown also in Fig. \ref{figura0} where we plot the time evolution of the magnetization $\vec f(t) = \int d{\bf r} \Psi_{0\alpha}^*({\bf r},t) {\bf F}_{\alpha \beta} \Psi_{0\alpha}^*({\bf r},t)$ with full 3D GP, see Ref. \cite{Zamora_Zamora_2019} for details of our methods, and $\vec f(t) =  \zeta_{\alpha}^*(t) {\bf F}_{\alpha \beta} \zeta_{\alpha}^*(t)$ with the SMA-GP equations. The agreement is essentially perfect. The unphysical feature of the SMA-GP equations, as well as of the full 3D GP, is that they are incapable of showing decoherence effects. This is not surprising since GP equations assume that there is a single macroscopic wavefunction for the ground state even in the presence of dynamical effects. 

The above results motivate, and partially justify, a radical single-mode approximation (SMA) that assumes that the field operator can be written as $\hat \psi_\alpha ({\bf r}) \approx \Psi_{0}({\bf r}) \hat  a_{\alpha}$, such that the espatial part can be integrated out and we can deal with the spin part in full. That is, the many-body aspects of the spin part can be fully taken into account. This will lead, as is the purpose of this paper, to show quantifiable aspects of intrinsic decoherence, as explained above, and the recurrence or revivals of the predictable oscillations for initial coherent states. The SMA approximation leads to a seemingly simple Hamiltonian for an $F=1$ SBEC, \cite{xue2018universal,law1998quantum,pu1999spin,Sarlo_2013}
\begin{equation}
\label{Ham}
\hat {H}=p \hat {f}_z+ q \hat {Q} +\eta \hat{f}^2. 
\end{equation}
Within SMA, this Hamiltonian differs from that of Eq. (\ref{Htot}) by a term proportional to the number operator $\hat N$, which commutes with $\hat H$. The first and second terms represent the linear and quadratic Zeeman contributions, $p \sim \tilde p B$, $q \sim \tilde q B^2$, while the third one is the spin-mixing interaction, $\eta \sim (c_2-c_0)$ such that $\eta > 0$ is polar and $\eta < 0$ ferromagnetic \cite{ho1998spinor}. As described in the Introduction our interest here is the ferromagnetic case. From now on, our goal is the study of Hamiltonian $\hat H$ and its ensuing dynamics and, therefore, we shall vary all three parameters independently in order to elucidate their role in the dynamics. These parameters can certainly be tuned experimentally. The above operators in $\hat H$ are given in terms of the creation and annihilation operators of bosonic atoms in the $z$-direction spin states $+1$, $0$ and $-1$, in obvious notation, 
\begin{eqnarray}
\hat f_z & = & \hat{a}_{+1}^{\dagger}\hat{a}_{+1} - \hat{a}_{-1}^{\dagger}\hat{a}_{-1}\nonumber \\
\hat Q & = &  \hat{a}_{+1}^{\dagger}\hat{a}_{+1} + \hat{a}_{-1}^{\dagger}\hat{a}_{-1} \nonumber \\
\hat f^2 & = & \hat {T}^\dagger+\hat {T} +(2 \hat{a}_0^\dagger \hat{a}_0-1) \hat{Q}+ \hat f_z^2
\end{eqnarray}
where
\begin{equation}
\hat{T}=2 \hat{a}_{-1}^{\dagger}\hat{a}_{+1}^{\dagger}\hat{a}_0\hat{a}_0 .
\end{equation}
Certainly $\hat{\vec{f}}=(\hat{f}_x,\hat{f}_y,\hat{f}_z)$ is the one-body spin or magnetization vector operator, with
\begin{eqnarray}
\hat f_x & = & \frac{1}{\sqrt{2}}\left(\hat{a}_{0}^{\dagger}(\hat{a}_{+1} + \hat{a}_{-1}) + (\hat{a}_{+1}^\dagger + \hat{a}_{-1}^\dagger)\hat{a}_{0}\right) \nonumber \\
\hat f_y &=& \frac{i}{\sqrt{2}} \left(\hat{a}_{0}^{\dagger}(\hat{a}_{+1} - \hat{a}_{-1}) - (\hat{a}_{+1}^\dagger -\hat{a}_{-1}^\dagger )\hat{a}_{0}\right)\label{fxy}
\end{eqnarray}
and $\hat{f}^2=\hat f_x^2+\hat f_y^2+\hat f_z^2$, and whose expectation value is the magnetization or spin texture. Introducing the spin state number operators $\hat n_{\sigma} = \hat a_\sigma^\dagger \hat a_\sigma$, one can also write $\hat f_z = \hat n_{+1} - \hat n_{-1}$ and $\hat Q = \hat n_{+1} + \hat n_{-1}$, forms that can be useful in interpreting our results below.

\begin{figure}[htbp]
\centering
\includegraphics[width=.45\textwidth]{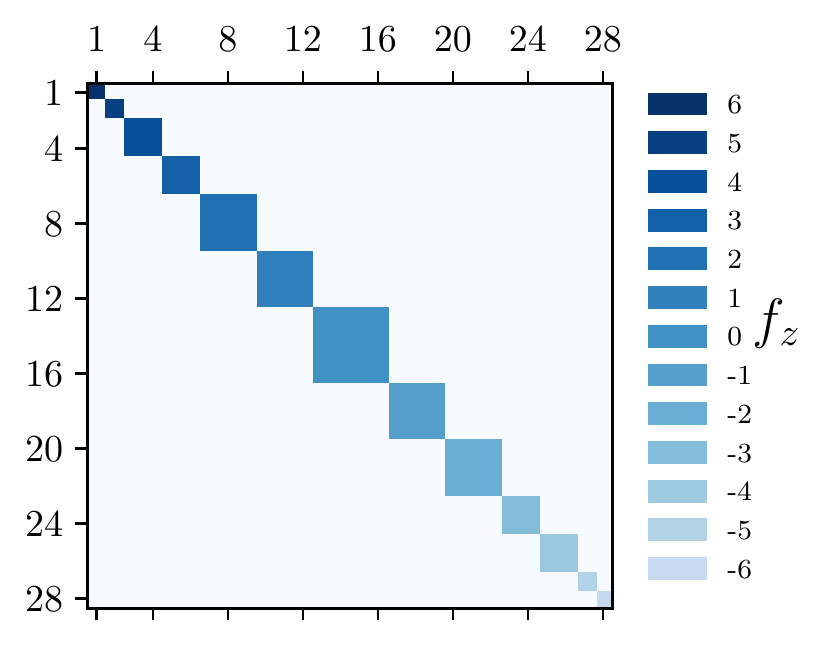}
\caption{Hamiltonian structure for N = 6 particles where the Hilbert space size is 28, each blue square represents a block of magnetization $M$. The intensity of color blue and the size of the blocks depend of the value of $M$.}
\label{figura1}
\end{figure}

\begin{figure}[h]
\includegraphics[width=.32\textwidth]{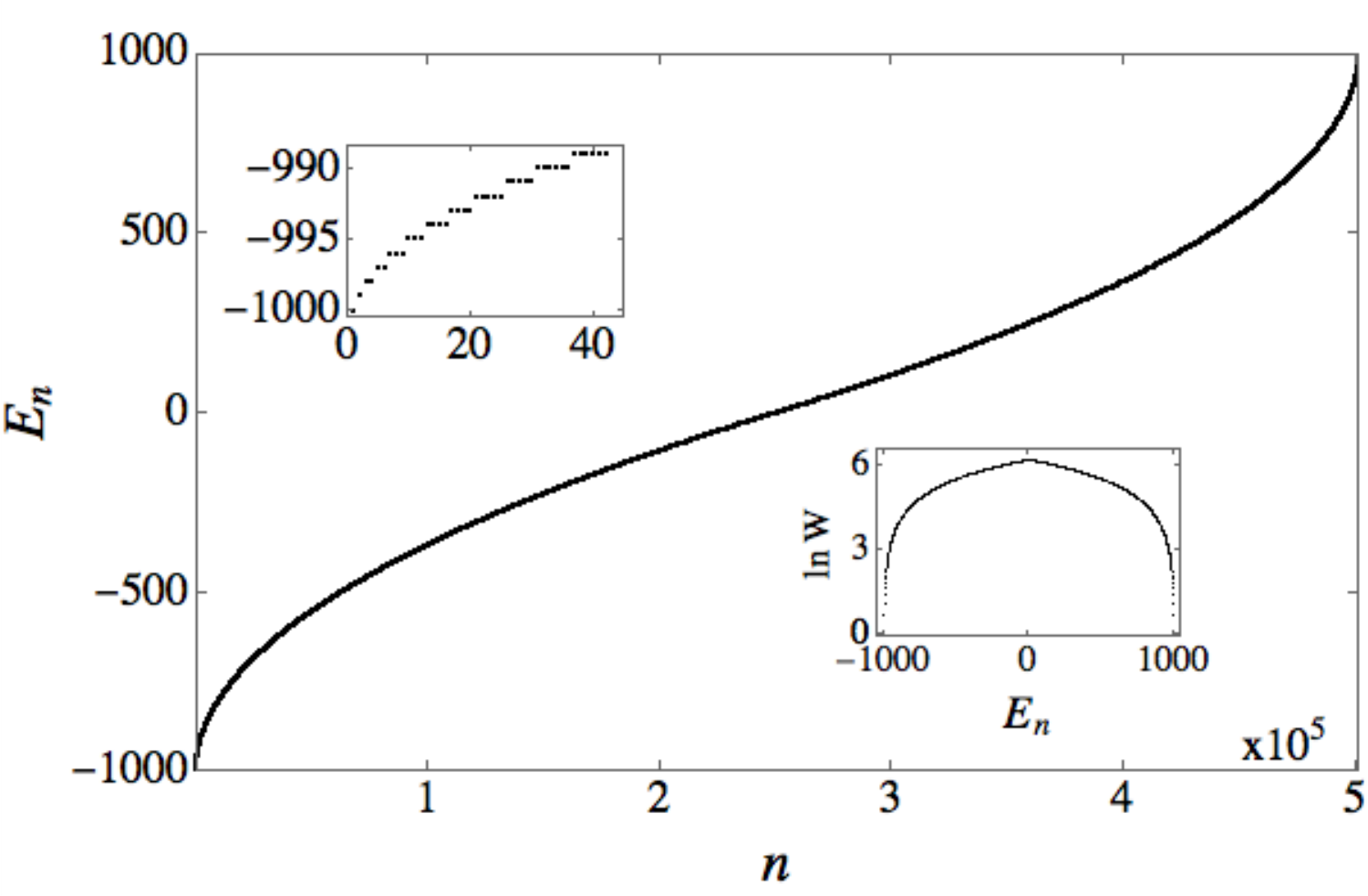}
\includegraphics[width=.32\textwidth]{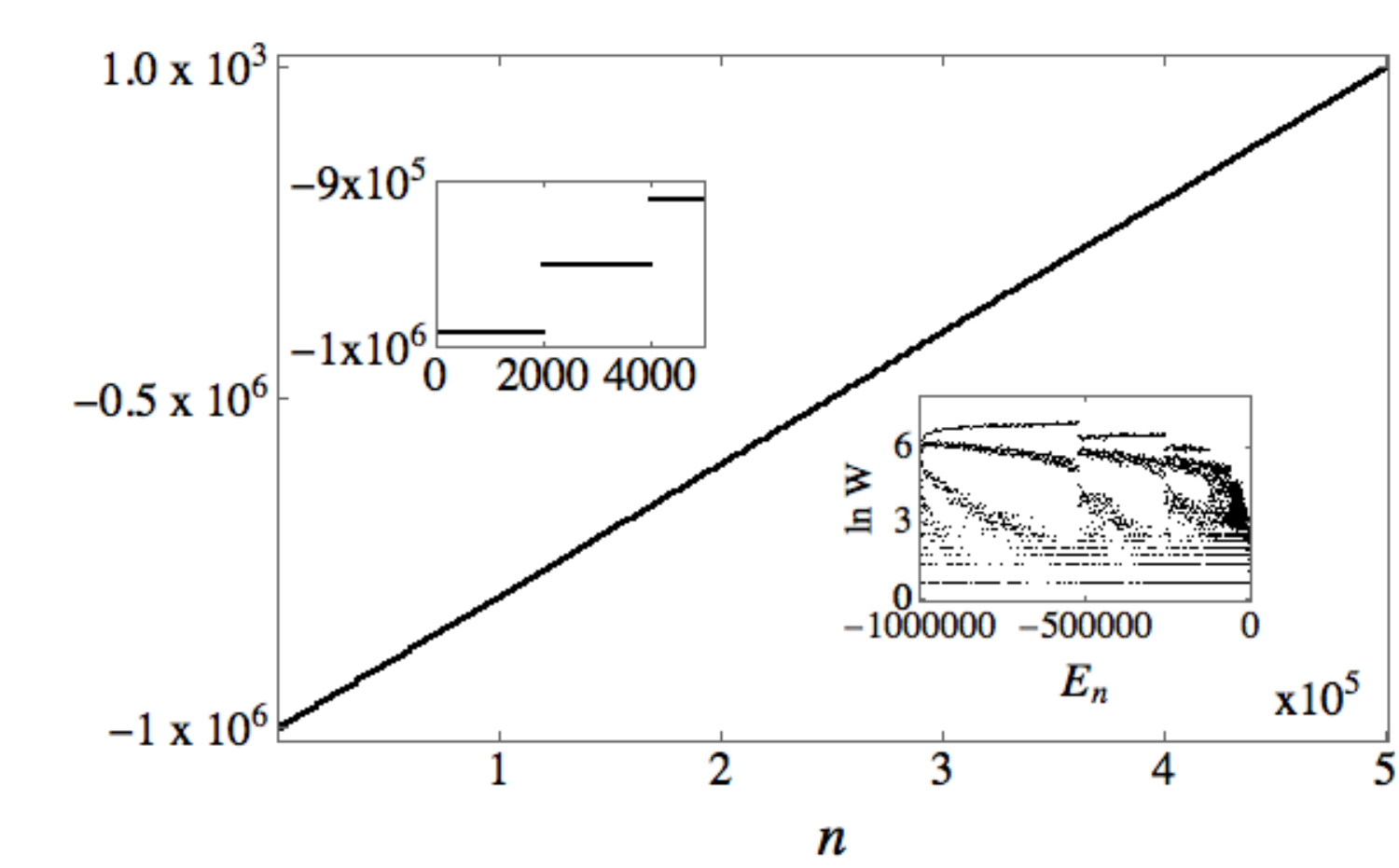}
\includegraphics[width=.32\textwidth]{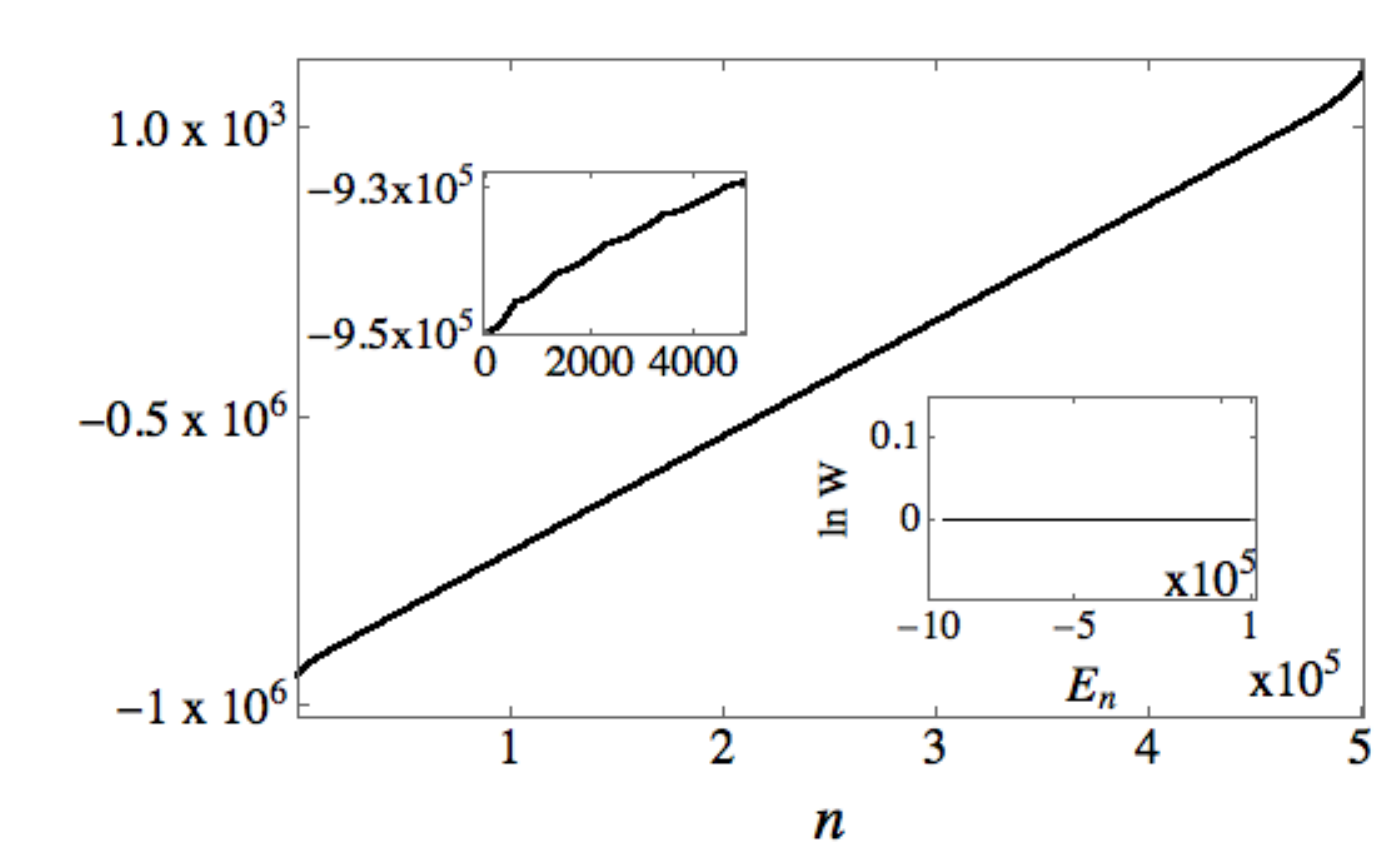}
\caption{Energy spectrum $E_n$ and its degeneracy $\ln W$ (right inset), for $N=10^3$ particles and (a) $p= 1$, $q = 0$, and $\eta = 0$; (b) $p = 0$, $q = 0$ and $\eta = -1$; and (c) $p = 1$, $q = 100$ and $\eta = -1$. In the left inset we show a detail of the energy spectrum.}
\label{Fig2}
\end{figure}

For a given number of atoms $N$
the size of the Hilbert space is $\Omega = (N + 1)(N + 2)/2$ and, therefore,
the size of Hamiltonian given by equation (\ref{Ham}) scales as $ \sim N^2 \times N^2$. In order to find the time evolution of the system we have to diagonalize the Hamiltonian, a difficult task that can be eased by exploiting its symmetries.
Obviously, the total number operator $\hat N = \hat n_{+1} + \hat n_0+\hat n_{-1}$ commutes with the Hamiltonian $\hat H$. In addition, due to its Lie structure, it is easy to show that $\left[\hat{f}_z, \hat{H} \right]=0$. Hence, instead of using the ``natural'' basis of number occupation $|n_{+1},n_0,n_{-1}\rangle$, in an obvious notation, one finds a better alternative to use $|M,n_0\rangle$, with $M = n_{+1}-n_{-1}$ the eigenvalues of $\hat f_z$, with values $-N, -N+1, \dots, N-1, N$. Atom number conservation $N = n_{+1} + n_0+ n_{-1}$ yields the third quantum number, obviated in the state labels. 
In this basis, the Hamiltonian is block diagonal in $M$ and the matrix elements show simple expressions,
\begin{equation}
\langle M^{\prime}, n_0^{\prime}|\hat{f}_z |M, n_0 \rangle=M \delta_{M^\prime M} \delta_{n_0^{\prime}n_0} \, ,
\label{melefz}
\end{equation}

\begin{equation}
\langle M^{\prime}, n_0^{\prime}|\hat Q |M, n_0 \rangle=(N-n_0) \delta_{M^{\prime} M} \delta_{n_0^{\prime} n_0} \, ,
\label{melefQ}
\end{equation}
\begin{eqnarray}
\langle M^{\prime}, n_0^{\prime}|\hat f^2 |M, n_0 \rangle&=& 2\sqrt{(n_{-1}+1)(n_{-1}+1)n_0 (n_0-1)} \delta_{M^{\prime} M} \delta_{n_0^{\prime} (n_0-2)} \nonumber \\
&+&2 \sqrt{n_{+1}n_{-1}(n_0+1)(n_0+2)} \delta_{M^{\prime} M} \delta_{n_0^{\prime} (n_0+2)} \nonumber \\
&+&\left(2 n_0-1\right)\left(N-n_0 \right) \delta_{M^{\prime} M} \delta_{n_0^{\prime} (n_0+2)} + M^2  \delta_{M^{\prime} M} \delta_{n_0^{\prime} n_0}\, ,
\label{melef2}
\end{eqnarray}
where $n_{+1}=(N+M-n_0)/2$ and $n_{-1}=(N-M-n_0)/2$. We note that the matrix elements in Eqs. (\ref{melefz})-(\ref{melef2}) do not link blocks with different values of $M$, however, they connect different matrix elements with jumps of 2 for $n_0$. Then, the Hamiltonian $H$ can be diagonalized and written as,
\begin{equation}
\hat H | M, m_M \rangle = E_{M,m_M} | M, m_M \rangle \label{basdiag}
\end{equation}
where, for each block $M$, $m_M = 1, 2, \dots, m_M^{\rm max}$, and $m_M^{\rm max} = (N-|M|+1)/2$ or $m_M^{\rm max} = (N-|M|+2)/2$ if $M$ is odd or even, respectively. This analysis can also be applied to SBEC with $F>1$ \cite{koashi2000exact,santos2006spin}. The block-diagonal structure of the Hamiltonian given by Eq. (\ref{Ham}) allows not only to very accurately calculate the energy eigenvectors and eigenvalues for values up to $N \sim 10^4$, as given by Eq. (\ref{basdiag}), but also to implement full quantum evolution of any initial quantum state for arbitrary values of time. For this, we perform numerically exact diagonalization block by block using pyCUDA linear algebra routines, in a server with 8 CPU Core, 128 GB RAM and with a graphic card Nvidia Tesla C2075. Since this diagonalization can be obtained for a wide variety of parameters, the ensuing time evolutions we obtain do not suffer from accumulation of errors, thus maintaining the same numerical precision at any time step. Depending on the value of $N$ and the needed time steps, the full evolution of a given initial state elapsed between few GPU seconds up to several hours. As an example easy to visualize, in Fig. \ref{figura1} we display the Hamiltonian structure for $N=6$ particles with a Hilbert space size of 28, each blue square representing a block of magnetization $M$, whose size is given below. The intensity of color blue and the size of the blocks depends on the value of $M$. Although we do not address it here, we show in Fig. \ref{Fig2} the spectra and its degeneracy for different values of $p$, $q$ and $\eta$ for $N=10^3$, to illustrate its richness. All the figures below are in dimensionless units. Since the three parameters of Hamiltonian $\hat H$ all have units of energy, for ease of varying all parameters without worrying on units adjustments in different cases, we assume a unit of energy $\epsilon_0$, such that the three parameters are adimensionalized with it and time is adimensionalized as $\tau = \epsilon_0 t/\hbar$.

Despite the fact that $\hat f_z$ and $\hat Q$ commute, the interesting and rich behavior of this model arises because of the non-conmutativity of the quadratic Zeeman term $\sim \hat Q$ and the spin interaction $\sim \hat f^2$. Further, in the absence of the quadratic Zeeman interaction, $q = 0$, the Hamiltonian can be analytically diagonalized both for $F = 1$ \cite{raghavan2000properties} and $F=2$\cite{koashi2000exact}. The presence of the quadratic term, $q \ne 0$, breaks the axisymmetry \cite{xue2018universal} and this system becomes an excellent one to study Quantum Phase Transition \cite{bookjans2011quantum}, quench-dynamical behaviors \cite{daug2018classification}, Quantum Kibble-Zurek Mechanism \cite{anquez2016quantum}, spin fragmentation \cite{de2013spin} among other mechanisms. On the contrary, if there is no atomic interactions, $\eta = 0$, the problem is also trivial and right away diagonal as given by Eqs. (\ref{melefz})-(\ref{melef2}). Hence, reiterating, the presence of the atomic interactions is mediated by the presence of the Zeeman quadratic term.

\section{Time evolution of the one-body density matrix in coherent states}

With the full quantum diagonalization described above we can calculate the time evolution of arbitrary initial states $|\Psi_0\rangle$. Here, we will concentrate in a family of initial coherent states, introduced below. For this purpose we briefly mention that, as expected, the time evolution can also be performed by blocks in the following manner. First, since the unitary propagator operator can be written as,
\begin{equation}
\mathbf{U}(t,t_0)= \sum_{M=-N}^N\sum_{m=1}^{m_M{\rm max}} e^{-\frac{i}{\hbar}  E_{M,m_M} (t-t_0) } |M,m_M\rangle \langle M,m_M|\>, \label{U}
\end{equation}
one just needs the overlaps $\langle M,m_M|\Psi_0\rangle$ to find the time evolution of any
$|\Psi_0\rangle$. However, since many of the usual physical quantities are typically one- or two-body operators, we shall devote our attention to the one-body density matrix, which allows for calculating all the statistical properties of all one-body operators, such as the matrices $\hat f_x$, $\hat f_y$, and $\hat f_z$. For this we note that since any one-body operator can be written as,
\begin{equation}
\hat {\cal O}^{(1)} = \sum_{kj} {\cal O}_{ij} \hat a_k^\dagger \hat a_j
\end{equation}
with $k$ and $j$ taking the values $+1$, $0$ and $-1$ and ${\cal O}_{ij}$ complex numbers, the expectation value of any operator $\hat {\cal O}^{(1)}$ requires the knowledge of the expectation values of the operators $a_k^\dagger a_j$ for all values of $kj$. These expectation values are those of the one-body density matrix. Explicitly, for a given initial state $|\Psi_0\rangle$, the one-body density matrix for all times is given by,
\begin{equation}
\rho_{jk}(t)  =  \frac{1}{N}\langle \Psi_0 |\mathbf{U}^\dagger(t,0) \hat a_k^\dagger \hat a_j \mathbf{U}(t,0) |\Psi_0\rangle ,\label{rhokj0}
\end{equation}
where the $1/N$ factor is introduced such that the trace of the reduced density is always unity. This yields in turn that the expectation values of spin operators are also bounded by one. Although we do not exploit here, it is worth mentioning that if we limit ourselves to one-body properties, one could also use the properties of the Gell-Mann spin 1 matrices \cite{GellMann}. A very interesting and useful property of the one-body density matrix given above, in the representation of the $\hat a_\alpha$, $\alpha = -1, 0, +1$ corresponding to the $z-$direction spin basis, is that the one-body reduced density matrix is diagonal in such a basis,
\begin{equation}
\rho_{jk}^{(s)} = \frac{1}{N}\langle M,m_M | \hat a_k^\dagger \hat a_k | M, m\rangle\delta_{jk} \>.\label{statmatrix}
\end{equation}
While we have not proved this result, we have extensively verified it and we believe it follows from the commutation of $\hat f_z$ with the Hamiltonian. The fact that the true stationary density matrix $\rho_{jk}^{(s)}$ is diagonal in this basis, will greatly facilitate the elucidation of the (quasi) stationary states reached in the time evolution of an initial coherent state.

As we can accurately calculate the matrix $\rho_{kj}(t)$ for any time using Eq.(\ref{rhokj0}), we now turn our attention to the initial  set of coherent states,\cite{kajtoch2016spin}
\begin{equation}
|\theta,\varphi \rangle=\frac{1}{\sqrt{N!}} \left(\zeta_1 \hat{a}^{\dagger}_{1}+\zeta_2\hat{a}^{\dagger}_{2}+\zeta_3 \hat{a}^{\dagger}_{3} \right)^N |0 \rangle \, ,\label{coherent}
\end{equation}
where $\zeta_1=e^{-i\varphi} \cos\left( \frac{\theta}{2}\right)^2$, $\zeta_2=\frac{\sin \theta}{\sqrt{2}}$ and $\zeta_3= e^{i\varphi} \sin\left( \frac{\theta}{2}\right)^2$, with $\theta$ and $\varphi$ the usual angles of the unit sphere, and $|0\rangle$ denoting the vacuum state with no particles. Alternatively, a coherent state can also be written as,
\begin{equation}
|\theta,\varphi \rangle =  e^{-i \varphi \hat f_z} e^{-i\theta \hat f_y} |N,0,0\rangle \>. \label{coherente2}
\end{equation}
A very important property to take into account is that these coherent states are eigenstates of $\hat f^2$, that is, $\hat f^2 |\theta,\varphi \rangle = N(N-1) |\theta,\varphi \rangle$. We point out that the distribution of energy eigenstates in an arbitrary 
coherent state involve, in general, several if not many blocks of different values of the quantum number $M$.
The main features of the time evolution of these states, as we amply discuss and show below, is that the elements of the one-body density matrix in these states 
 show an initial oscillation that suffers intrinsic decoherence followed by a stationary state and revivals at later times, with this behavior being repeated ad infinitum. It is evident that both the decoherence and recurrence times depend on the  $p$, $q$ and $\eta$ parameters as well as on the initial state $(\theta,\phi)$, but an important issue is its dependence on $N$. As mentioned in the Introduction we have indentified that the dependence on $N$ is very different in two opposite limits $N|\eta|/q \ll 1$ and $q/N|\eta| \ll 1$, evidently called weak and strong interacting regimes.

Since one can show that the set of operator $\hat f_z$, $\hat Q$ and $\hat f^2$ are not part of a Lie algebra,  the finding of an analytic expression for the reduced density matrix appears as a very difficult task. Nonetheless, the main contribution of this article, is to show that the time evolution of the SBEC one-body properties can be summarized quite precisely with explicit (semi) analytic expressions for the time evolution of the density matrix elements $\rho_{kj}(t)$, given in Eq. (\ref{rhokj0}), for arbitrary initial coherent state such as Eq.(\ref{coherent}), in the weak and strong limits.  These expressions, as well as their validity limits, are found in a heuristic manner based on a very large number of precise numerical evaluations of time evolutions for a wide variety of values of the Hamiltonian parameters and for a collection of different initial coherent states. 

In order to introduce our expressions for the one-body density matrix in the following section, we first discuss preliminary exact results. Note that the one-body density matrix $\rho_{jk}(t)$, given by (\ref{rhokj0}), can be first expressed as,
\begin{equation}
\rho_{jk}(t) = \langle \theta |  \mathbf{U}_{QI}^\dagger(t,0) a_k^\dagger a_j \mathbf{U}_{QI}(t,0)  | \theta \rangle e^{i (j-k) (p t/\hbar+\varphi)} \label{rhokj2}
\end{equation}
where
\begin{equation}
 \mathbf{U}_{QI}(t,0) = e^{-i \left(q \hat Q + \eta \hat f^2\right)t/\hbar} \label{UQI}
\end{equation}
and $|\theta\rangle \equiv |\theta,0\rangle$ is the coherent state for any value of $\theta$ and $\varphi = 0$. We see that for any initial state the role of the $p\hat f_z$ term and the angle $\varphi$ simply amount to a phase factor, while the whole dynamics is ruled by the quadratic Zeeman $q\hat Q$ and the interaction $\eta \hat f^2$ terms. Whether one term or the other dominates depends on the relative values of the strength parameters $q$ and $\eta$. We further note that if $q = 0$, the interactions play no role since the coherent states are eigenstates of $\hat f^2$. Therefore, in order to observe the effect of both terms, both $\eta$ and $q$ must be nonzero. 

\begin{figure}[h]
\includegraphics[width=0.33\textwidth]{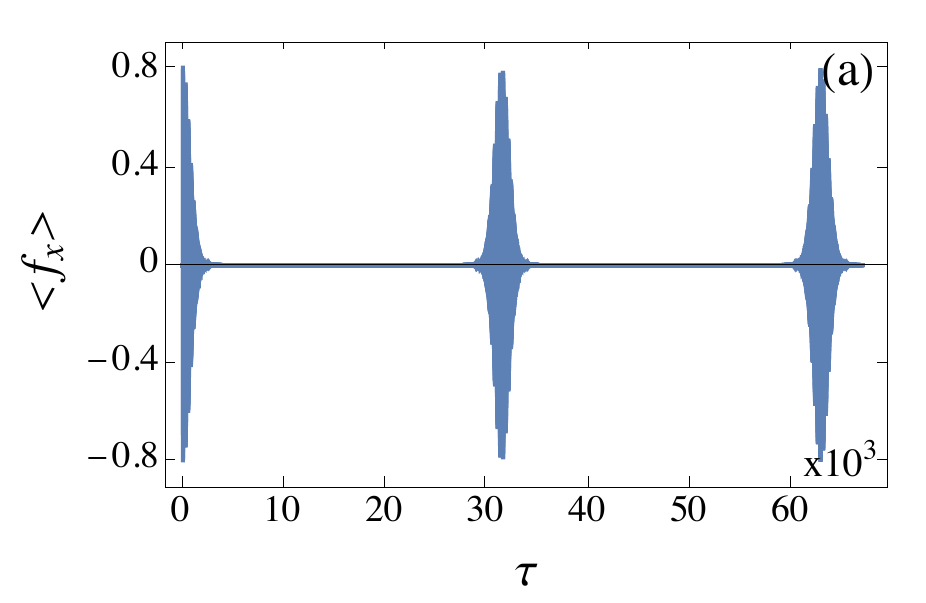}
\includegraphics[width=0.33\textwidth]{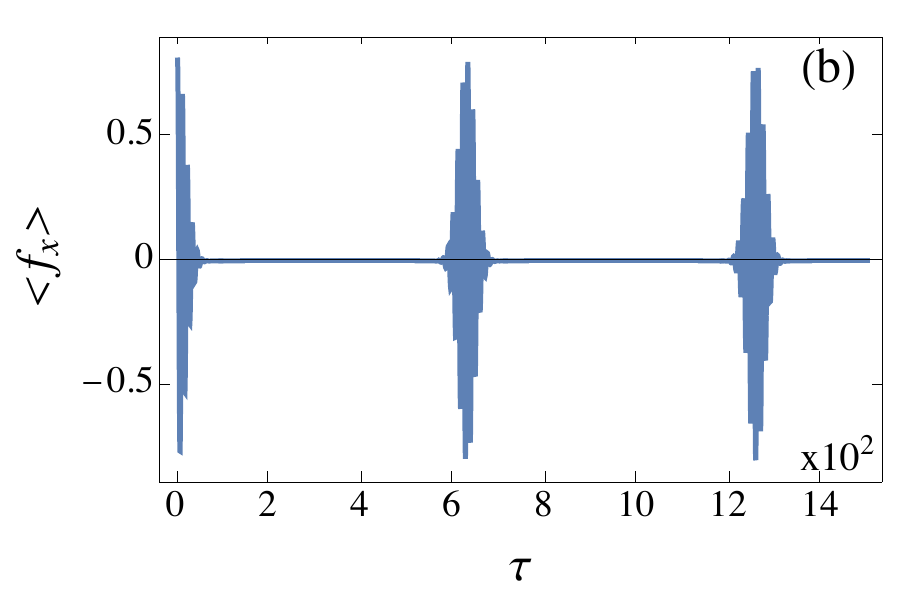}
\includegraphics[width=0.33\textwidth]{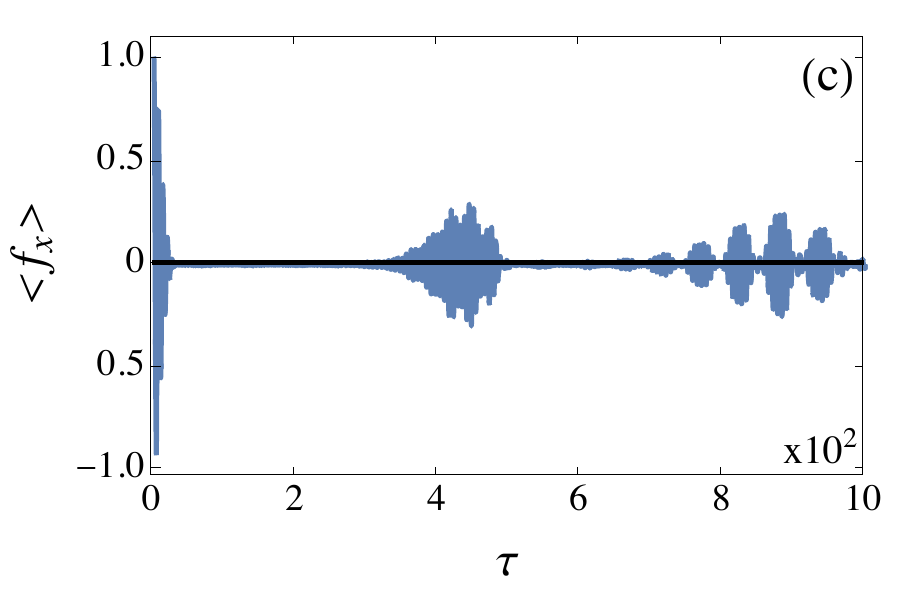}
\caption{Time evolution of the expectation value $\langle \hat f_x \rangle$ illustrating the sequence of decoherences and recurrences; (a) weak interaction $N|\eta|/q \ll 1$ ($q = 1$, $\eta = -10^{-4}$, $\theta = 3\pi/10$); (b) strong interaction $q/N|\eta| \ll 1$ ($q = 1$, $\eta = -10^4$, $\theta = 3\pi/10$), and (c) crossover $q/N|\eta| \sim 1$ ($q = 1$, $\eta = -10^{-2}$, $\theta = \pi/2$). In all cases  $N = 100$.}
\label{figmuestra}
\end{figure}

As an illustration of typical time evolution of coherent states, in Fig. \ref{figmuestra} we show sequences of decoherences and recurrences in the weak $N|\eta|/q \ll 1$, strong $q/N|\eta| \ll 1$ and crossover $N|\eta|/q \sim 1$ regimes. We have found that in the extreme cases the recurrences appear at periodic intervals, thus making feasible their prediction, while in the crossover, as the two effects contribute similarly, the sequences can be very irregular and we make no attempt to analyze them here.

\section{Decoherence and recurrences in the strong and weak interacting regimes}

To proceed with the finding of the weak and strong interacting extremes, we note that the density matrix, Eq. (\ref{rhokj2}), can be written in two alternative forms,
\begin{equation}
\rho_{jk}(t)  =  \langle \theta |e_T^{-\frac{i}{\hbar}\eta \int_0^\tau \hat V_I(\tau) d\tau}\>a_k^\dagger a_j \>e_T^{\frac{i}{\hbar}\eta \int_0^\tau \hat V_I(\tau) d\tau}| \theta\rangle
e^{i((k-j)pt/\hbar+\varphi)} e^{i \delta_{k, j \pm 1} (\delta_{j,2}-\delta_{k,2}) qt/\hbar} \label{rhoeta}
\end{equation}
or 
\begin{equation}
\rho_{jk}(t)  =  \langle \theta |e_T^{-\frac{i}{\hbar}q \int_0^\tau \hat V_Q(\tau) d\tau}\> a_k^\dagger a_j \> e_T^{\frac{i}{\hbar}q \int_0^\tau \hat V_Q(\tau) d\tau}| \theta\rangle
e^{i((k-j)pt/\hbar+\varphi)} \>.\label{rhoq}
\end{equation}
where
\begin{eqnarray}
\hat V_I(\tau) &=& e^{iq \hat Q \tau/\hbar} \hat f^2 e^{-iq \hat Q \tau/\hbar} \nonumber \\
\hat V_Q(\tau) &=& e^{i\eta \hat f^2 \tau/\hbar} \hat Q e^{-i\eta \hat f^2 \tau/\hbar} \>,
\end{eqnarray}
and $e_T$ stands for the time-ordered exponential. The first form, Eq.(\ref{rhoeta}), is appropriate for the study of the weak interaction $N|\eta|/q \ll1$, while the second one, Eq.(\ref{rhoq}), for the strong one $q/N|\eta| \ll 1$. Note that the exact limits $\eta = 0$ and $q = 0$ are explicitly recovered in Eqs. (\ref{rhoeta}) and (\ref{rhoq}).

In addition to the analysis of a large number of numerical calculations of the time evolution of coherent states, varying all the relevant parameters, we also used as insight the analytical solution of the non-linear single-mode Hamiltonian $\hat h = -\mu \hat a^\dagger \hat a+ u \hat a^\dagger \hat a^\dagger \hat a \hat a$, see Refs. \cite{Imamoglu} and \cite{plimak2006quantum}, that for  usual harmonic oscillator coherent states \cite{Cohen-Tannoudji} show  collapses whose envelope is of a gaussian shape followed by revivals, with characteristic times that depend exclusively on the state and the Hamiltonian parameters. Furthermore, the recurrences are observed to appear at periodic times. Based on this and on the numerical evidence, we proceed now to present the heuristic (semi) analytic forms of the density matrix in the two extremes.

\subsection{Strong interaction regime $q/N|\eta| \ll 1$} 

In this regime the leading terms depend mostly on the quadratic Zeeman coupling $q$ and the one-body reduced density matrix can be very precisely fitted by the following form,
\begin{equation}
\rho_{jk}(t) \approx  \rho_{jk}(0) e^{\frac{N}{2} \sin^2 \theta \left[ \cos \left(\frac{q (k-j)}{\hbar N} t\right)-1\right]} e^{i \cos \theta (k-j) q t/\hbar} e^{i (k-j) p t/\hbar} \>\>\>\>j \ne k \>.\label{rhoqaprox}
\end{equation}
with $\rho_{jk}(0)$ given by,
\begin{equation}
\rho_{jk}(0) = 
\left(
\begin{array}{ccc}
\cos^4 \frac{\theta}{2} & \frac{1}{\sqrt{2}}e^{i \varphi} \sin\theta \cos^2\frac{\theta}{2} & \frac{1}{4} e^{2 i \varphi}\sin^2\theta  \\
\frac{1}{\sqrt{2}}e^{- i \varphi} \sin\theta \cos^2\frac{\theta}{2} &  \frac{1}{2} \sin^2\theta & \frac{1}{\sqrt{2}}e^{-i \varphi}\sin\theta \sin^2\frac{\theta}{2}  \\
\frac{1}{4} e^{2i \varphi}\sin^2\theta   & \frac{1}{\sqrt{2}}e^{i \varphi}\sin\theta\sin^2\frac{\theta}{2}  & \sin^4 \frac{\theta}{2}
\end{array}\right) \>.\label{rhokjini}\,
\end{equation}
which was obtained using the lie algebra method \cite{Sandoval}

From Eq. (\ref{rhoqaprox}, it can immediately be seen that the structure of periodic recurrences with decoherences in relatively shorter times than the former is given by the real exponential term  
$\exp(\frac{N}{2} \sin^2 \theta \left[ \cos \left(\frac{q (j-k)}{\hbar N} t\right)-1\right])$, indicating that at the times $\tau_n = n \tau_{rec}^{(q)}$, $n = 0, 1, 2, \dots$, periodic recurrences occur,
\begin{equation}
\tau_{rec}^{(q)}=\frac{2 N \pi}{(j-k)q } \>.\label{trecq}
\end{equation}
We label this as the strong-interaction recurrence time. The strong-interaction decoherence time can be readily find by expanding the previous exponential at short times, yielding a gaussian function of the form $e^{-t^2/2\tau_{dec}^2}$, from which we identify,
\begin{equation}
\tau_{dec}^{(q)}=\sqrt{\frac{N}{2}} \>\frac{\hbar}{q (j-k)\sin\theta } \>.\label{tdecq}
\end{equation}
Note that both times $\tau_{rec}^{(q)}$ and $\tau_{dec}^{(q)}$ grow as does the number $N$ of atoms that, as we discuss in the next section, appear as an ``expected'' feature of thermodynamic behavior in the limit of large number of atoms. 

 The diagonal terms $\rho_{jj}(t)$, for $j = -1,0,+1$, are essentially constant equal to their initial values at $t = 0$. The precision of our calculations allows to set,
\begin{equation}
\rho_{jj}(t) \approx  \rho_{jj}(0) + a_{jj}(\theta) e^{\frac{N}{2} \sin^2 \theta \left[ \cos \left( \frac{4q}{\hbar N} t\right)-1\right]} \>.\label{rhoqaproxdiag}
\end{equation}
with $a_{jj}(\theta)$ a term that it is at most of the order of $10^{-9}$, compared with the initial value $\rho_{kj}(0)$, for $N \lesssim 10^3$. The decoherence time scales as $\tau_{dec}^{(q)}\sim \sqrt{N}$ and the recurrence one $\tau_{rec}^{(q)}\sim N$, in agreement with this regime, see Eq. (\ref{tdecq}).

\begin{figure}[htbp]
\centering
\includegraphics[width=.45\textwidth]{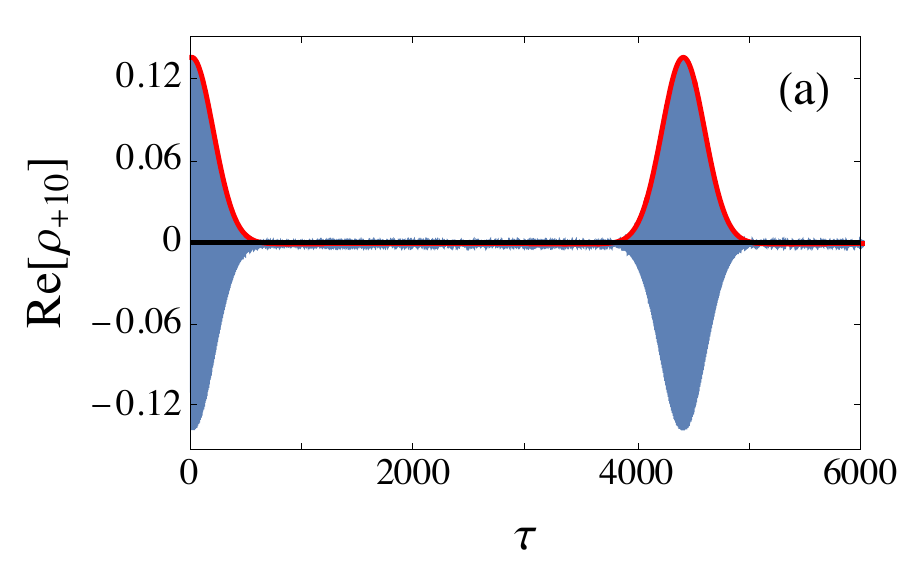}
\includegraphics[width=.45\textwidth]{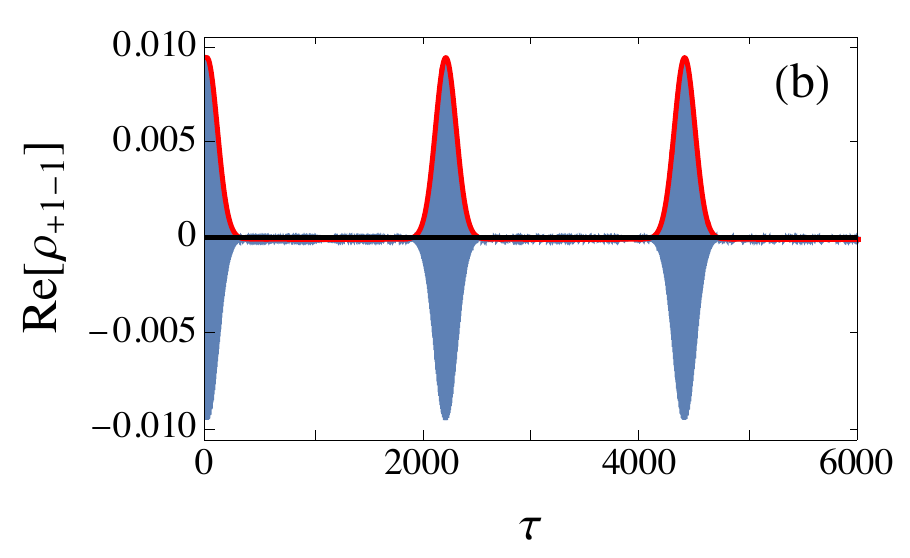}
\includegraphics[width=.45\textwidth]{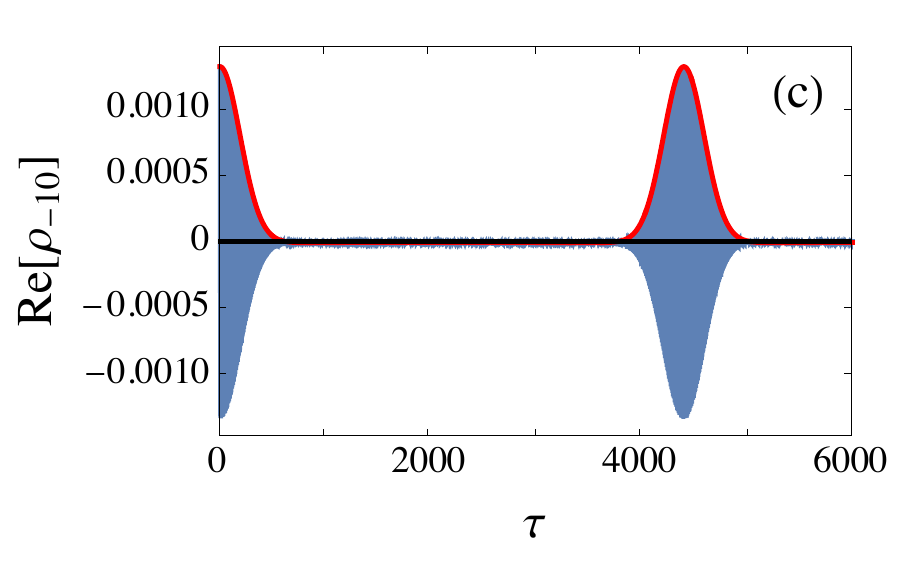}
\includegraphics[width=.45\textwidth]{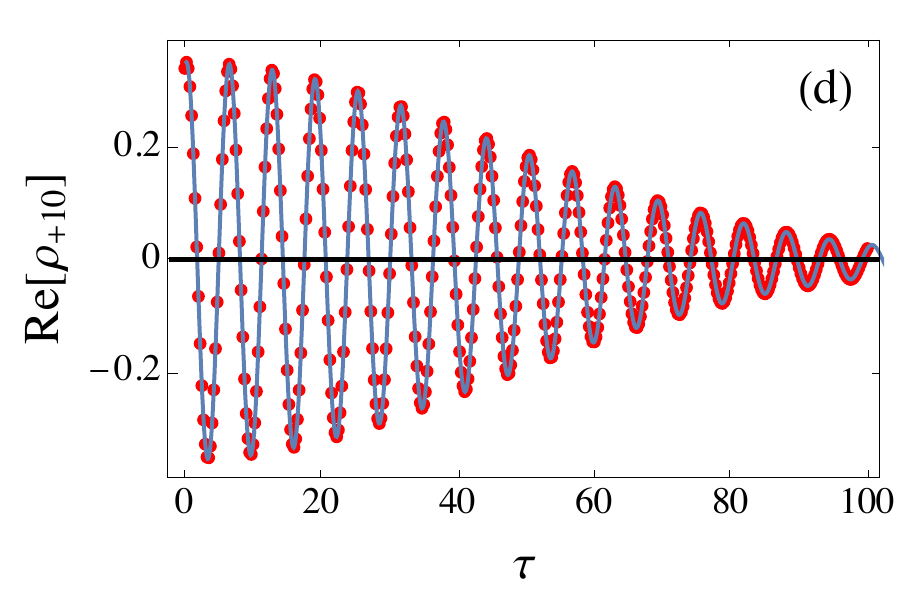}
\caption{Real part of elements (a) $\rho_{+1,0}$, (b)  $\rho_{+1,-1}$ and (c) $\rho_{-1,0}$ of the density matrix as a function of time, for $N = 700$, $q = 1$, $p = 1$ and $\eta = - 30000$, for a time period longer than $(0,2\pi\hbar N/ q)$. In continuous (red) lines we show the overlap of the recurrences predicted by Eq (\ref{rhoqaprox}). In panel (d) we show 
an example of the agreement of the oscillations predicted by Eq. (\ref{rhoqaproxdiag}), red dots, with the full quantum calculation, continuous blue line, for the real part of $\rho_{+1,0}$;  $N = 1000$, $q = 1$, $p = 1$ and $\eta = 30000$.}
\label{Figstrong}
\end{figure}

In Fig. \ref{Figstrong} we show an example of the evolution of  the density matrix in the strong interacting regime $q/N|\eta| \ll 1$ ( $q = 1$, $p = 1$ and $\eta = -30000$ for $N = 700$). This illustrates both the typical behavior of decoherences and recurrences and the agreement with the heuristic fitting given by Eq. (\ref{rhoqaprox}). In panels (a), (b) and (c) we show the evolution of the real part of the off-diagonal elements of the density matrix that show the sequence of decoherences followed by stationary states and recurrences. In panel (d) we show the agreement with the predicted oscillations in terms of the $q$ and $p$ parameters.

\begin{figure}[h]
\begin{center}
\includegraphics[width=.7\textwidth]{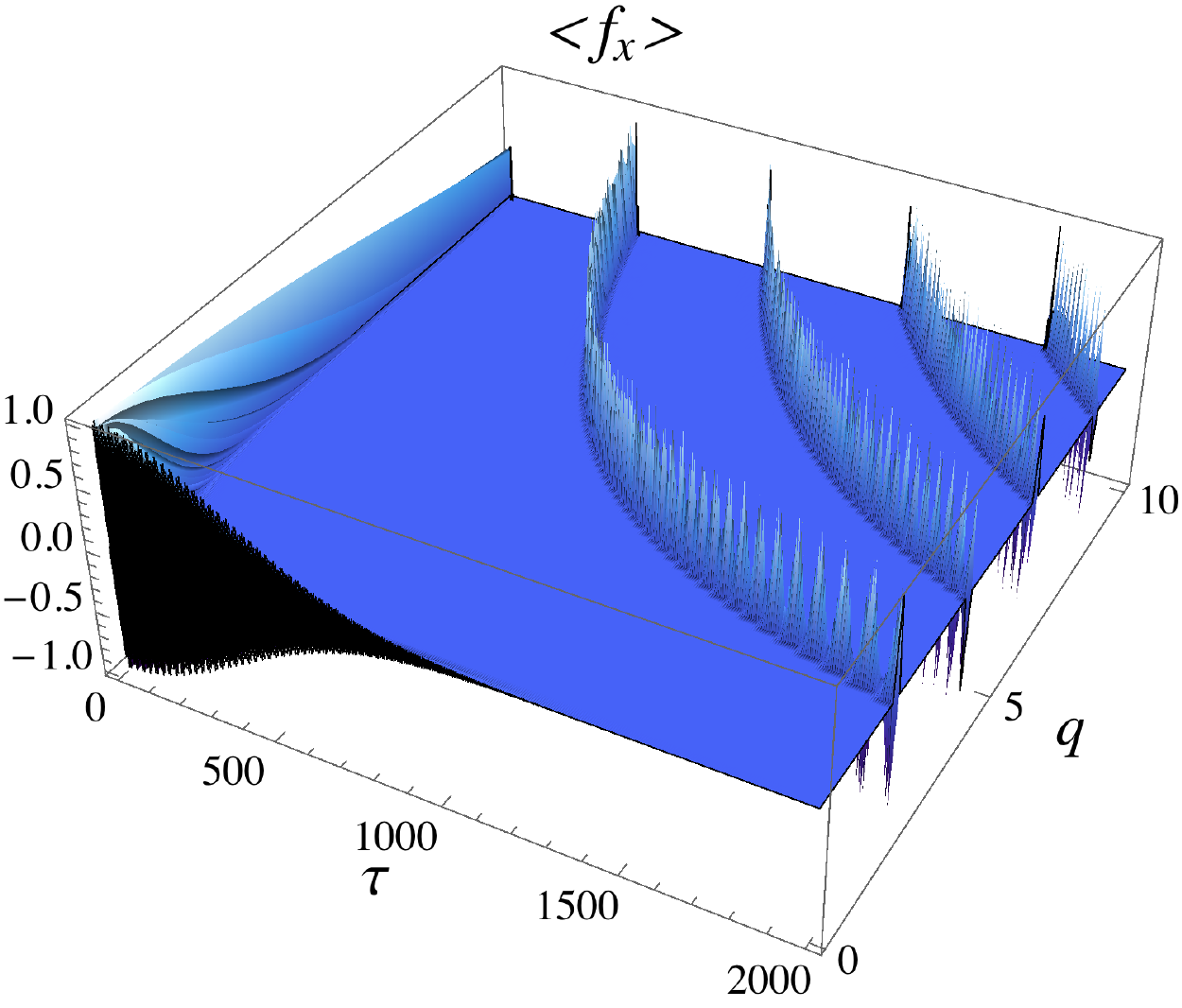}
\caption{Evolution in time of the expectation value $\langle \hat f_x \rangle$ as a function of time $\tau$ and the non-linear Zeeman strength $q$, for $p = 1$ and $\eta = -30000$. Note that if $q \ll 1$, the recurrences appear more separated.}
\label{muchas}
\end{center}
\end{figure}

 As shown in Fig. \ref{muchas}, and confirmed by Eq. (\ref{rhoqaprox}), the behavior of one-body observables, such as $\langle \hat f_x\rangle$, show a very detailed and predicted structure of decoherences and recurrences, with a stationary state for a very large parameter space. As seen in this and in all the above figures and expressions, the one-body density matrix of the stationary state is diagonal in the spin basis $(-1,0,+1)$ along the $z-$direction. This physically means that while the expectation value of the $z-$component of the spin vector $\hat {\vec f}$, as well as the non-linear Zeeman term $\hat Q$, remain essentially constant, the $x$ and $y$ spin components decohere to zero.
Note from the figure that if $q \ll 1$, but $\eta$ finite, the recurrences can be separated by very long times that increase as $N$ grows, thus yielding a true stationary state in the thermodynamic limit. To enquire into the nature of the stationary state, as illustrated in Figs. \ref{Figstrong} and \ref{muchas} and in accord with Eqs. (\ref{rhoqaprox}) and (\ref{rhoqaproxdiag}), we observe that the (quasi) stationary one-body density matrix becomes diagonal with their diagonal terms numerically very close to the initial ones. As suggested by ETH, we compute the average energy $\overline E = \langle \theta, \varphi |\hat H| \theta, \varphi \rangle$ and find the closest eigenenergy $E_{M,m_M}$ of the system to such an average. We have found in all studied cases for this regime that the corresponding stationary density matrix $\rho_{jk}$ equals the 
 density matrix $\rho_{jk}^{(s)}$ of the corresponding energy eigenstate $|M,m_M\rangle$, see Eq. (\ref{statmatrix}). Although we do not claim that the stationary state is a thermal, this result agrees with the consequences of the Eigenstate Thermalization Hypothesis. That is, all the one-body properties in the (quasi) stationary states are the same as if the system were in a true, exact, stationary energy eigenstate with the same eigenvalue as the average energy of the time-evolving system. As depicted in Fig. \ref{Figstrong}, the recurrences become more separated as $N$ increases, thus indicating that for a very large system, the quasi stationary state cannot be distinguished from a true eigenstate, as far as measurable few-body properties are concerned. This is also along the explanation of thermalization in isolated many-body systems in statistical physics \cite{LL}.

\subsection{Weak interaction regime $N|\eta|/q \ll 1$}

\begin{figure}[h]
\includegraphics[width=0.4\textwidth]{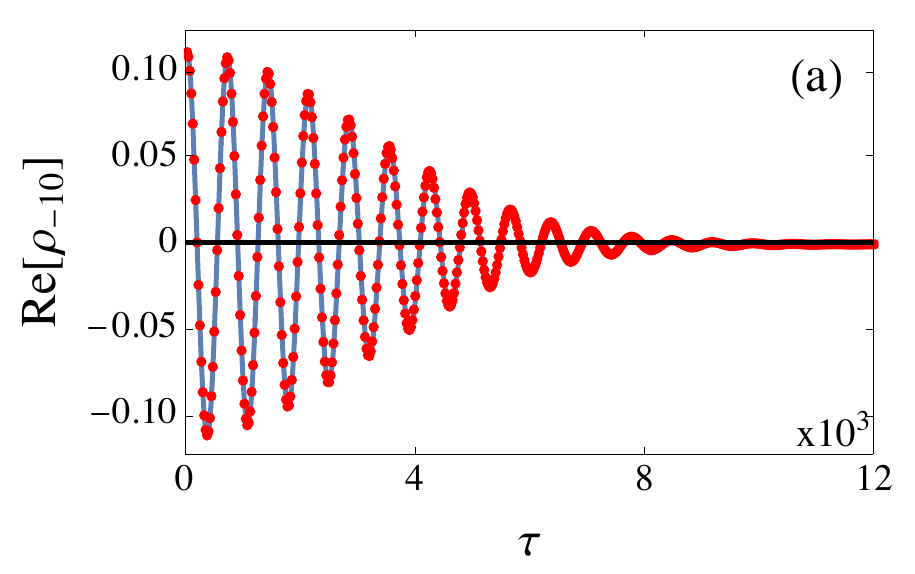}
\includegraphics[width=0.4\textwidth]{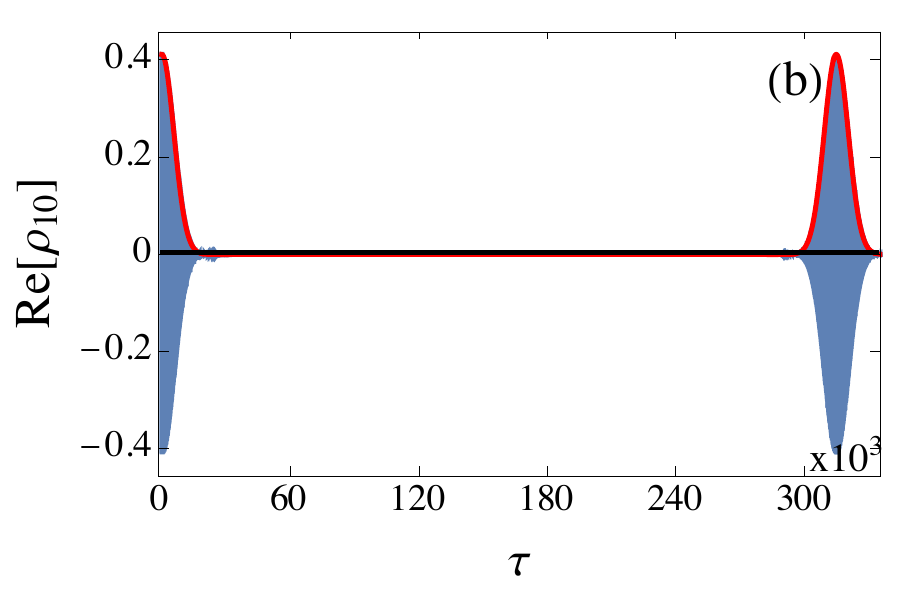}
\includegraphics[width=0.4\textwidth]{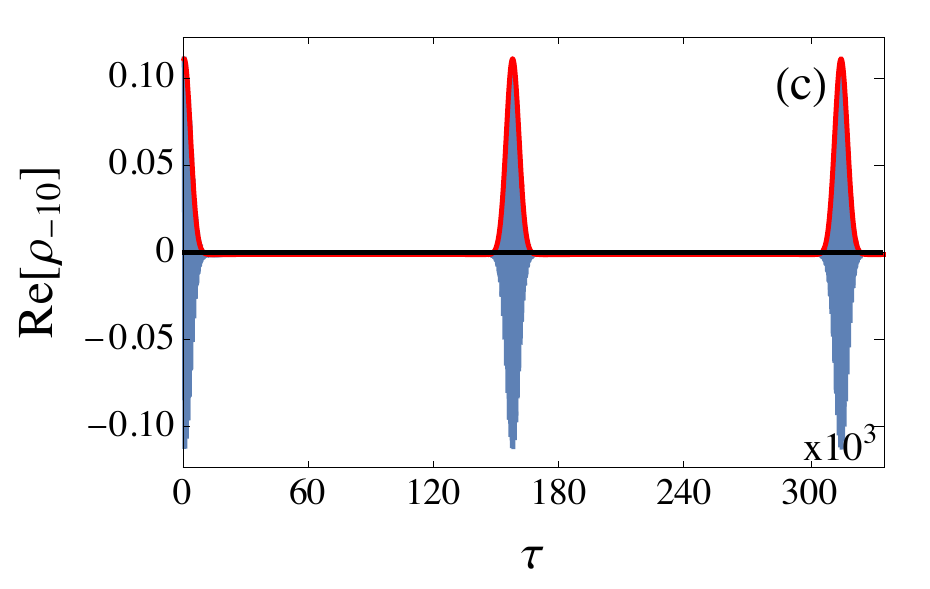}
\includegraphics[width=0.4\textwidth]{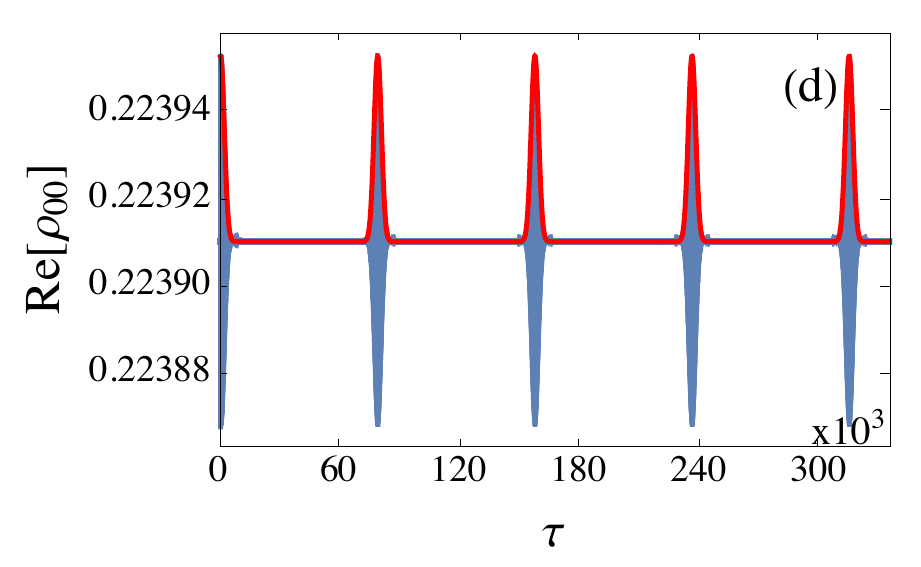}
\caption{Real part of different elements of the density matrix as a function of time. (a) Comparison of full quantum calculation (blue solid line) with heuristic fit (red dots), Eq.(\ref{rhojk}), of  $\rho_{+1,-1}$. Recurrences of real parts of  (b) $\rho_{+1,0}$, (c)   $\rho_{+1,-1}$ and  $\rho_{0,0}$, within the time period $(0,2\pi\hbar/\eta)$; full quantum calculation (blue solid line) and overlap of  heuristic fit (red line), Eq.(\ref{rhojk}). The (dimensionless) parameters are $N = 300$, $\eta = -10^{-5}$; $q = 7$, $\theta = 7 \pi/30$, $p = 0$ and $\varphi = 0$.}
\label{Figro13}
\end{figure}

Figure \ref{Figro13} shows typical behavior of elements of the one-body density matrix in the weak interaction regime, for a particular case, see figure caption. We find that all essential features of the time evolution of the one-body reduced density matrix can be fitted quite well by the following explicit expressions.

\begin{figure}[h]
\includegraphics[width=.3\textwidth]{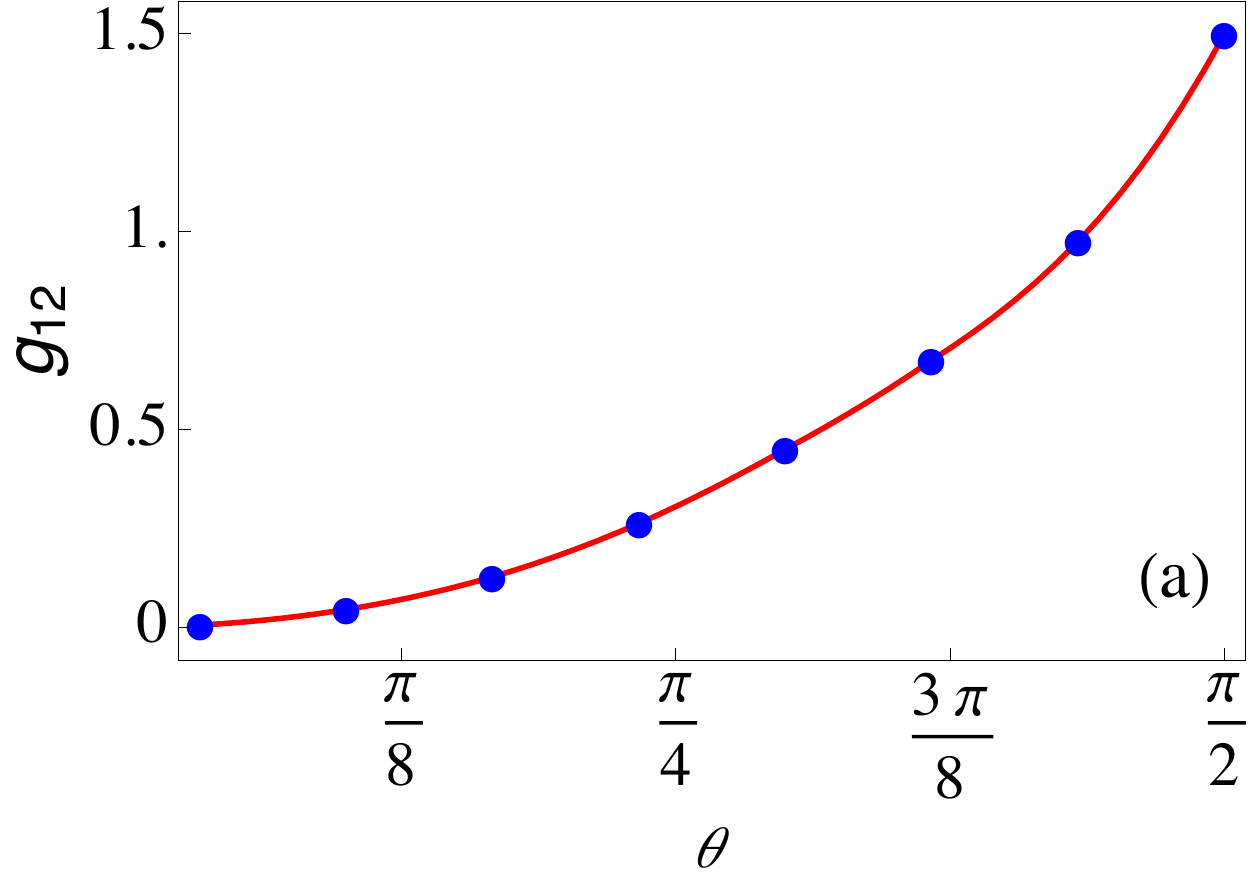}
\includegraphics[width=.3\textwidth]{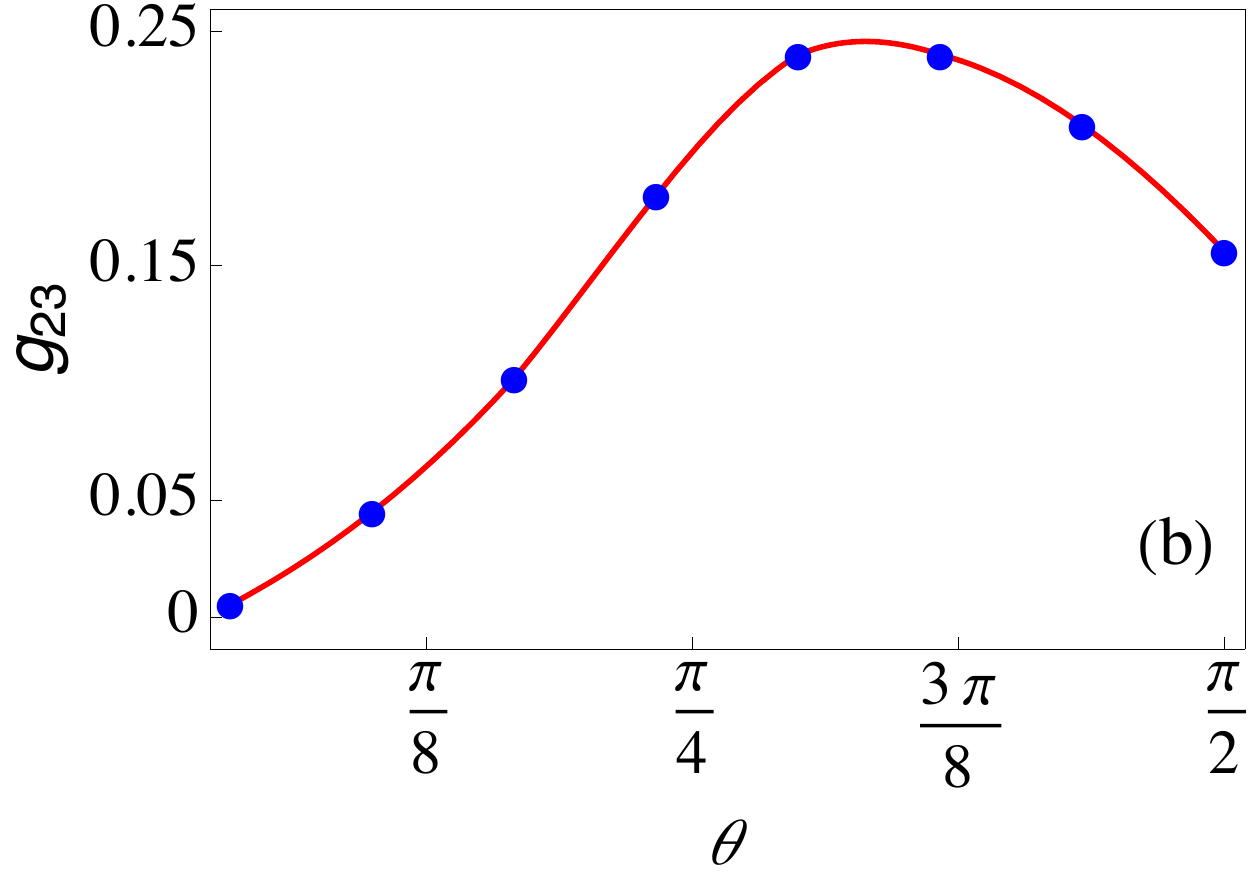}
\includegraphics[width=.3\textwidth]{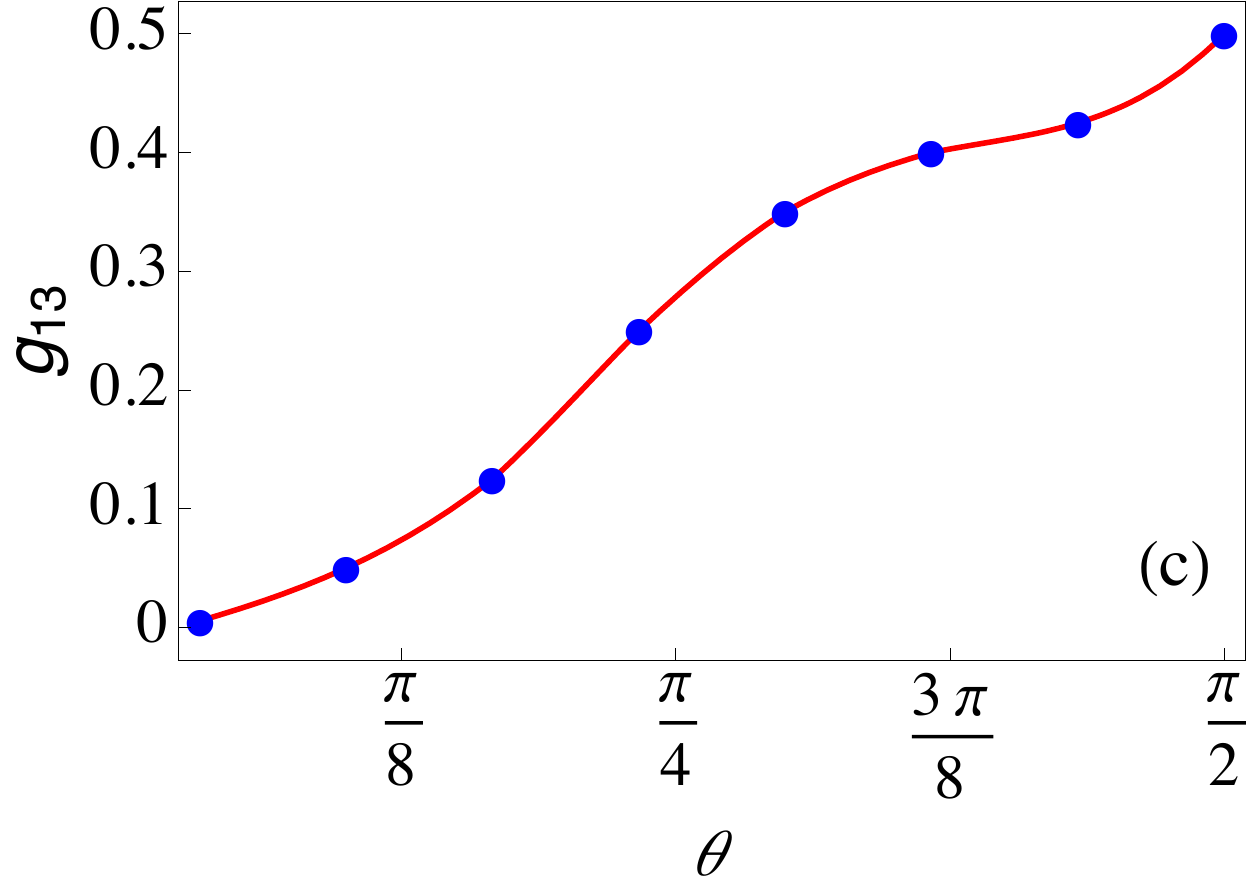}
\caption{Coefficient $g_{jk}(\theta)$ as a function of $\theta$, in Eq. (\ref{Fjk}) . (a) $g_{12}(\theta)$. (b) $g_{23}(\theta)$. (c) $g_{13}(\theta)$. The dots are values numercially calculated and the continuous line is a spline interpolation. The parameters are $N = 300$, $\eta = 10^{-5}$; $q = 7$, $p = 0$ and $\varphi = 0$.}
\label{fit-off}
\end{figure}

\begin{figure}[h]
\includegraphics[width=.3\textwidth]{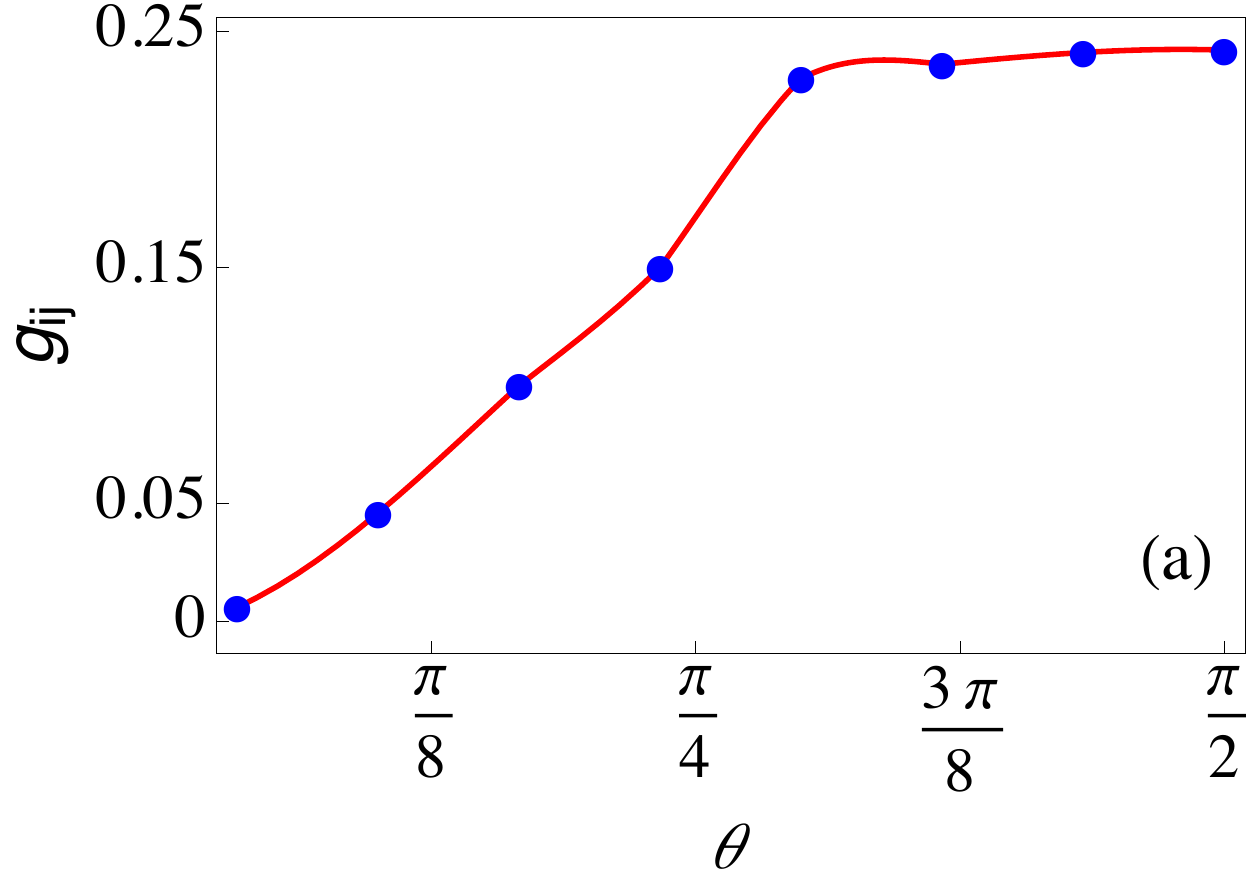}
\includegraphics[width=.3\textwidth]{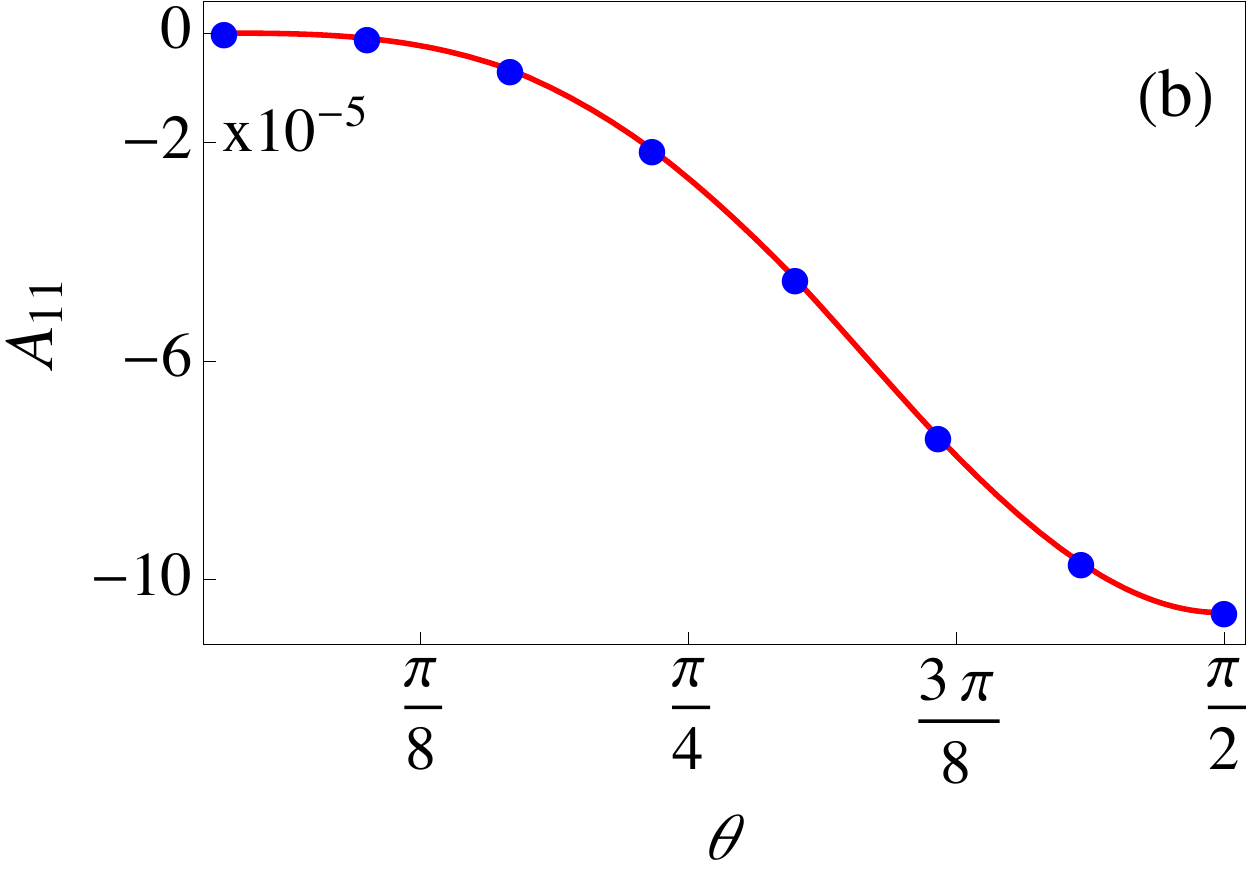}
\includegraphics[width=.3\textwidth]{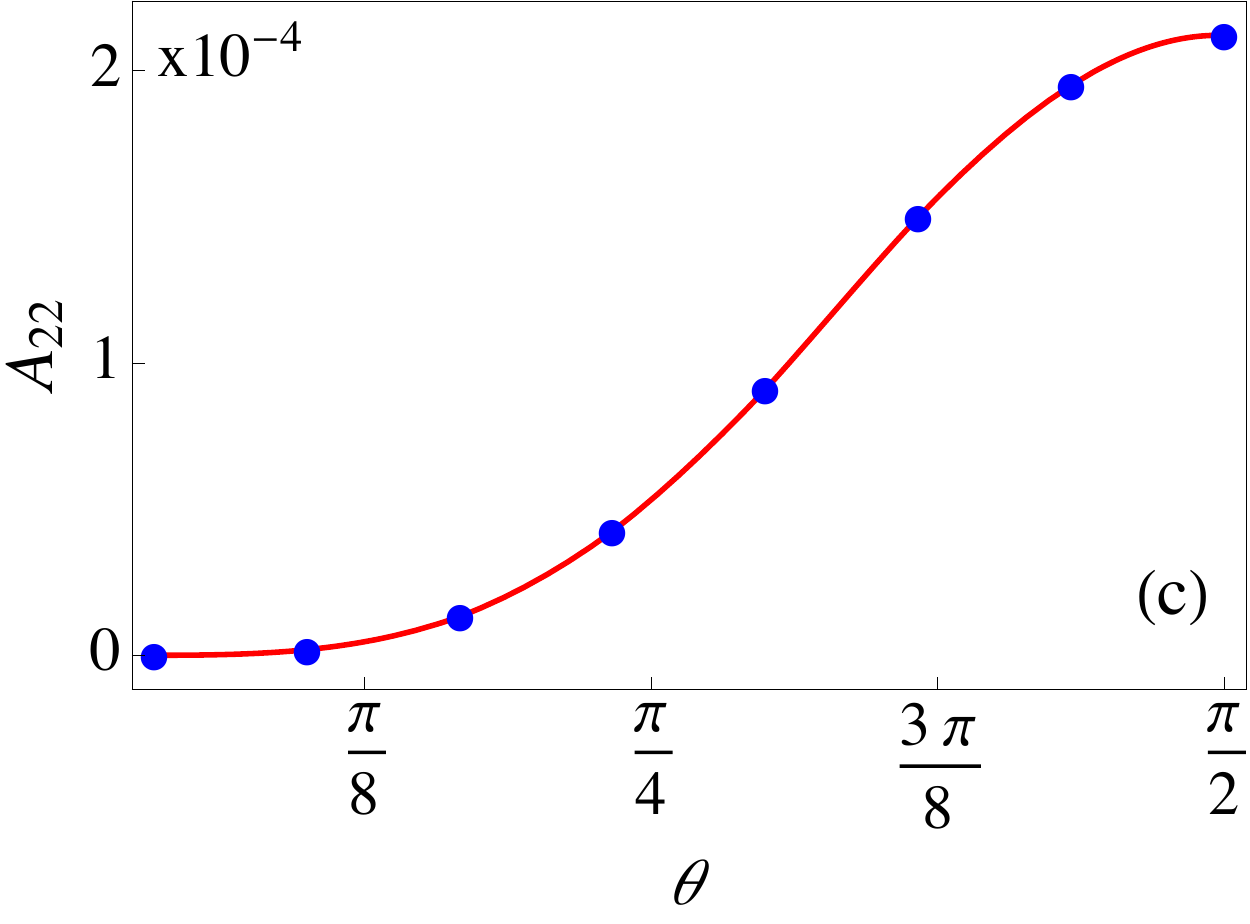}
\caption{Amplitude coefficient (a) $A_{11}(\theta)$ and (b) $A_{22}(\theta)$ as a function of $\theta$, in Eq. (\ref{rhojj}). Note that both are very small. (c) Coefficient $g_{11}(\theta) = g_{22}(\theta) = g_{33}(\theta)$as a function of $\theta$ in Eq. (\ref{Fjk}).  The dots are values numerically calculated and the continuous line is a spline interpolation. The parameters are $N = 300$, $\eta = 10^{-5}$; $q = 7$, $p = 0$ and $\varphi = 0$.}
\label{fit-diag}
\end{figure}

\begin{equation}
\rho_{jk}(t) \approx \rho_{jk}(0) F_{jk}(N,\eta,\theta,t) e^{-2i(j-k) \eta N \cos\theta t} e^{-i(j-k)pt}e^{i\delta_{j,k\pm1} (\delta_{k,0}-\delta_{j,0})qt} \>\>\>\>\>j \ne k \label{rhojk}
\end{equation}
\begin{equation}
\rho_{jj}(t) \approx \rho_{jj}(0)+ A_{jj}(\theta)\left(1- F_{jj}(N,\eta,\theta,t)\right) \cos \frac{qt}{2} \label{rhojj}
\end{equation}
where
\begin{equation}
F_{jk}(N,\eta,\theta,t) = e^{g_{jk}(\theta)N\left(\cos(2 f_{jk} \eta t)-1\right)} \>\>\>\>\textrm{for all}\> j,k \>.\label{Fjk}
\end{equation}
Above, $j,k = -1,0,+1$; $f_{ij}$ are symmetrical with $f_{1,0} = 1$, $f_{-1,1}=2$, $f_{-1,0} = 3$ and all diagonal equal, $f_{jj} =  4$.
We were not able to find analytic expressions for the coefficients $g_{jk}(\theta)$ and $A_{jj}(\theta)$ but they can be numerically fitted, as we show them in Figs.  \ref{fit-off} and \ref{fit-diag}. The diagonal terms $g_{jj}(\theta)$ are all equal, shown in Fig. \ref{fit-diag}. The initial condition is given again by the matrix in Eq. (\ref{rhokjini}).

\begin{figure}[h]
\includegraphics[width=.45\textwidth]{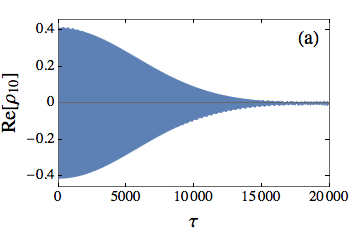}
\includegraphics[width=.45\textwidth]{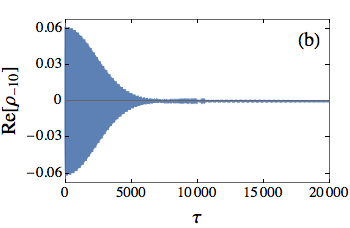}
\includegraphics[width=.45\textwidth]{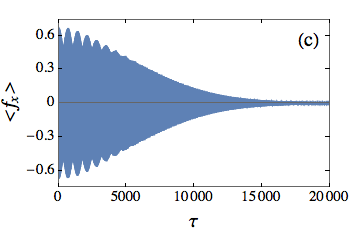}
\includegraphics[width=.45\textwidth]{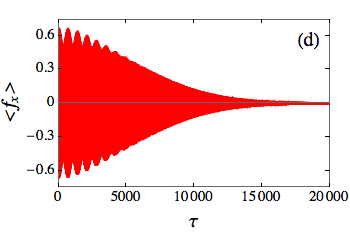}
\caption{(a) Real part of element $\rho_{+1,0}$ of the density matrix as a function of time. (b) Real part of element $\rho_{-1,0}$ of the density matrix as a function of time. (c) Expectation value of $\hat f_x$ as a function of time, full quantum calculation; (c) Expectation value of $\hat f_x$ as a function of time, heuristic fit, Eq. (\ref{rhojk}). The beating pattern in $\langle \hat f_x(t) \rangle$ is due to a very small difference of the high frequency oscillations of oscillations between $\rho_{+1,0}$ and $\rho_{-1,0}$, see Eq.(\ref{rhojk}). The parameters are $N = 300$, $\eta = -10^{-5}$; $q = 7$, $\theta = 7 \pi/30$, $p = 0$ and $\varphi = 0$.}
\label{Figrofx}
\end{figure}

Figs. \ref{Figro13} and \ref{Figrofx} show the density matrix $\rho_{jk}(t)$ and the expectation value of the operator $\hat f_x$ for a single example in the limit $N|\eta|/q \ll 1$ ($\eta = -10^{-5}$ and $q = 7$, for $N = 300$). These figures illustrate the excellent agreement of the heuristic expression given by Eq. (\ref{rhojk}) with the accurate numerical calculations. Fig. \ref{Figro13}(a) shows both cases for the initial decoherence, while Figs. \ref{Figro13}(b), (c) and (d) the agreement in predicting the occurrence of the recurrences, for $\rho_{-1,0}$, $\rho_{-1,1}$ and $\rho_{0,0}$. Fig. \ref{Figrofx} shows the behavior of the elements (a) $\rho_{1,0}$ and (b) $\rho_{-1,0}$. As can be seen in Eq. (\ref{rhojk}), these two elements oscillate with a very large frequency $q = 7$ and with a very small correction $\pm 2 N \eta \cos \theta$, each taking one of the signs. This tiny difference between the oscillations of $\rho_{1,0}$ and (b) $\rho_{-1,0}$ shows itself in the beating pattern of the expectation value of $\hat f_x$ (see Eq.(\ref{fxy})) shown in Fig. \ref{Figrofx}(c) and (d): the former is the exact numerical calculation and the latter the heuristic expression given by Eq. (\ref{rhojk}). We found a quite remarkable agreement  for any value of $\theta$.

The behavior of intrinsic decoherences, followed by quasi-stationary states to further revivals or recurrences in a periodic fashion, is essentially contained in the function $F_{jk}(N,\eta,\theta,t) \approx e^{g_{jk}(\theta)N\left(\cos(2 f_{jk} \eta t)-1\right)}$ within the density matrix, see Eq.(\ref{rhojk}). The function $F_{jk}$ indicates right aways that there are recurrences at periodic intervals, $\tau_n = n \tau_{rec}$ with $n = 1, 2, 3, \dots$, and
\begin{equation}
\tau_{rec}^{(\eta)} = \frac{2 \pi}{h_{j,k}|\eta|}\>, \label{treceta}
\end{equation}
that we label as the weak-interacting recurrence time; it does depend on the particular matrix element but the main point is its  inverse proportionality to $\eta$ and its independence on $N$. Then, at each recurrence, starting with the initial time $t = 0$, the evolution appears to decohere in a shorter time scale. This can be found by approximating the exponential for short times,
\begin{equation}
e^{g_{jk}(\theta)N\left(\cos(2 f_{jk} \eta t)-1\right)} \approx e^{- g_{jk}(\theta)N \left(2 f_{jk} \eta t\right)^2/2} \>,
\end{equation}
thus identifying the weak-interacting decoherence time, $e^{-t^2/2\tau_{dec}^2}$,
\begin{equation}
\tau_{dec}^{(\eta)}= \sqrt{\frac{1}{g_{jk}(\theta)N}} \>\frac{\hbar}{2h_{jk}|\eta|} \>.\label{tdeceta}
\end{equation}
In this case, in addition to the dependence on $\eta$, there appears a dependence on the number $N$ of atoms that is very different to the strong interaction case, see Eqs. (\ref{trecq}) and (\ref{tdecq}). This is further discussed in the last section.

With regard to the (quasi) stationary states reached between consecutive recurrences or revivals,  as shown in Figs. \ref{Figro13} and \ref{Figrofx} and verified in expressions given by Eqs. (\ref{rhojk})-(\ref{Fjk}), the off-diagonal terms of the reduced density matrix again vanish for time intervals during the (quasi) stationary state, with the diagonal terms becoming equal to the initial state. As in the strong interaction regime, since the energy eigenstates one-body density matrices, 
see Eq. (\ref{statmatrix}), are also diagonal in the same spin basis, this suggests the same test of comparing the density matrices in the stationary regime of the coherent states with that of a true stationary eigenstate with the same average energy. Interestingly, contrary to the strong interacting regime, in this case those do not agree. That is, the stationary state between recurrences does not seem to be a ``true'' stationary state, in the sense of ETH. This suggests, along the mentioned $N$ dependence of the decoherence and recurrence times, that the weak regime is not typical of thermodynamic states. Although the full analysis of these stationary states are out of the scope of the present article, this indicates that the distribution of energy eigenstates in coherent states may show a more complex structure in the weak than in the strong limit.

\section{Discussion and Final Remarks}

The agreement between the proposed heuristic expressions for the time evolution of the one-body density matrix of a $F=1$ SBEC, compared with full (numerically) exact full quantum calculation is certainly quite good, allowing us to validate our conclusion that those expressions are the leading non-trivial order, becoming better as $N$ grows. Once more, in both cases the general behavior shows periodic (intrinsic) decoherences of an oscillatory behavior, into stationary states, with periodic recurrences. However,  the precise nature of this phenomenon is very different in the two limits considered. The most notorious difference between the extreme regimes resides on the fact that the role of the number of particles $N$ is very different in each case. Note that in the weak side $\tau_{dec}^{(\eta)} \sim 1/\sqrt{N}$ and $\tau_{rec}^{(\eta)} \sim {\cal O}(1)$, while in the strong one $\tau_{dec}^{(q)} \sim \sqrt{N}$ and $\tau_{rec}^{(q)} \sim N$. It is of interest to mention that the non-linear single-mode approximation, as described in Refs. \cite{Imamoglu} and \cite{plimak2006quantum}, corresponds to the weak-interaction regime.  We find relevant to recall that the validity of thermodynamics is achieved for macroscopic bodies $N \gg 1$. In this limit, averages of extensive quantities scale as $N$ while their fluctuations as $\sqrt{N}$ such that the ratio of deviations from equilibrium values scale as $1/\sqrt{N}$, hence tending to zero as $N$ increases \cite{LL}. Moreover, the larger the system, the longer it takes to equilibrate in a stationary state and the recurrences to initial states also become apart for a longer time. In this sense, it appears that the strong interacting regime fulfills this limit. That is, $\tau_{dec}^{(q)} \sim \sqrt{N}$  and $\tau_{rec}^{(q)} \sim N$ appear  as reasonable results. It is the more appealing the fact that the ratio $\tau_{dec}/\tau_{rec} \sim 1/\sqrt{N}$, the mentioned typical condition for thermodynamic stability of a macroscopic system \cite{LL}. Thus, although both times diverge in the thermodynamic limit, in the appropriate scale the decoherence time tends to zero as $\sim 1/\sqrt{N}$ compared to the recurrence one. Therefore, we find very interesting to observe that while the decoherence and recurrence times in the {\it weak}-interaction case, $N|\eta|/q \ll 1$, do not follow the ``expected'' thermodynamic trend,  still we observe that the ratio $\tau_{dec}^{(\eta)}/\tau_{rec}^{(\eta)} \sim 1/\sqrt{N}$ scales appropriately. In any case however, although the role of the number of atoms $N$ is very different in each limit, still the ultimate responsible for the decoherence and recurrence phenomena is the interaction among the constituents of the body. As we have also seen, this
strong difference appears to be present in the structure of the stationary states attained between consecutive recurrences, having started in coherent states. As we have verified, those stationary states are indistinguishably from true stationary energy eigenstates. The result that this property does not hold for a weakly interacting SBEC, appears to be in agreement with the $N$ dependence of the decoherence and recurrence times mentioned above. The full elucidation of these differences certainly deserves a separate and detailed study. In the crossover regime $N|\eta| \sim q$ both the interaction $\eta$ and non-linear Zeeman $q$ contributions compete, their effect is intertwined and even though the system indeed shows decoherences and recurrences, the latter no longer occur at prescribed times depending on either $\sqrt{N}$ or $N$, see Fig. \ref{figmuestra} (c). 

To summarize, we first highlight that one can exactly (numerically) diagonalize the Hamiltonian of an interacting $F = 1$ SBEC in the SMA approximation for large number of atoms and that the method can be extended to larger spins $F > 1$. This allows us to probe the full quantum dynamics of any initial state. For the purposes of the present study, we have chosen here the relevant family of the coherent states. We have studied the corresponding reduced one-body density matrix and have found heuristic analytical expressions that fit remarkably well its dynamics in the weak $N|\eta|/q \ll 1$ and strong $N|\eta|/q \gg 1$ regimes, indicating the interplay among the non-linear Zeeman effect $\sim q$ and the pairwise spin interaction $\sim \eta$, mediated by many-body effects $\sim N$. Our expressions predict the decoherence time and the recurrence periods in terms of these quantities and become more precise as $N$ grows. Since the corresponding unitary propagator cannot be analytically found, due to its lack of closed Lie algebra, our results may indicate a path to find it in a series whose leading order term agrees with the heuristic expressions here found. A natural extension of the present study is the analysis of two-body correlations.  For this we need to calculate the reduced two-body density matrix, which with our method does not appear as a very difficult task. It would be very interesting to find how these correlations behave along the different regimes, to find out if the predicted decoherence and recurrence time hold and it would also serve to further enquire into the structure and properties of the stationary state between recurrences. 

\acknowledgments{J.C.S.S. acknowledges a postdoctoral scholarship at UNAM during the time this research was done.  This work was partially funded by grant IN108620 DGAPA (UNAM) and CONACYT 255573.}


\bibliography{biblio.bib}

\end{document}